\documentclass[prd,aps,a4paper,nofootinbib,eqsecnum,twocolumn]{revtex4}   %-1,showpacs,showkeys

\newif\ifusesec
\usesectrue  
   
\usepackage{graphicx} 
\usepackage{mathrsfs}
\usepackage{amsmath,amsfonts,amssymb}
\usepackage{multirow}
\usepackage{eufrak}

\newcommand{\beq}{\begin{equation}}
\newcommand{\eeq}{\end{equation}}
\newcommand{\vS}{{\bf S}}
\begin{document}

\title{Gravitational spin-orbit coupling in binary systems at the second post-Minkowskian approximation} 

\author{Donato \surname{Bini}$^{1}$}
\author{Thibault \surname{Damour}$^2$}

\affiliation{
$^1$Istituto per le Applicazioni del Calcolo ``M. Picone'', CNR, I-00185 Rome, Italy\\
$^2$Institut des Hautes Etudes Scientifiques, 91440 Bures-sur-Yvette, France
}

\date{\today}

\begin{abstract}
We compute  the rotations, during a scattering encounter, of the spins of two gravitationally interacting particles at second-order
in the gravitational constant (second post-Minkowskian order). Following a strategy introduced in  Phys. Rev. D {\bf 96}, 104038 (2017), we transcribe our
result into a correspondingly improved knowledge of the spin-orbit sector of the Effective One-Body (EOB) Hamiltonian description
of the dynamics of spinning binary systems. We indicate ways of resumming our results for defining improved versions
of spinning EOB codes which might help in providing a better analytical description of the dynamics of coalescing
spinning binary black holes.
\end{abstract}

\maketitle

\section{Introduction} \label{sec1}

The Effective One-Body (EOB) formalism \cite{Buonanno:1998gg,Buonanno:2000ef,Damour:2000we,Damour:2001tu} 
is a  theoretical framework which yields a resummed analytical  description of  the gravitational dynamics of two spinning bodies. 
It can be essentially thought as mapping the dynamics (in the center-of-mass frame) of  two bodies, of masses 
$m_1,m_2$ and spins ${\bf S}_1$, ${\bf S}_2$, onto a Hamiltonian description of the motion of a single body, of mass 
\beq
\mu\equiv\frac{m_1 m_2}{m_1+m_2}
\eeq
 and (suitably rescaled) spin   
 \beq \label{Ss}
 {\bf S}_* \equiv \frac{m_2}{m_1}{\bf S}_1
+ \frac{m_1}{m_2}{\bf S}_2,
\eeq
moving (modulo some higher-in-momenta corrections, and modified spin-orbit effects) in an effective metric 
that is a deformed version of a Kerr metric of mass 
\beq
M\equiv m_1+m_2,
\eeq
 and spin
${\bf S}_0 = {\bf S}+  {\bf S}_*$, where
\beq \label{S}
{\bf S} \equiv {\bf S}_1 +{\bf S}_2\,.
\eeq
The key deformation parameter entering EOB theory is the symmetric mass ratio
\beq \label{nu}
 \nu  \equiv \frac{\mu}{M} = \frac{m_1 m_2}{(m_1+m_2)^2} \, .
\eeq

Since its introduction about twenty years ago \cite{Buonanno:1998gg,Buonanno:2000ef,Damour:2000we,Damour:2001tu}
the EOB approach has been continuously improved by incorporating new relevant results obtained  through various formalisms, such as Post-Newtonian (PN) theory, Numerical Relativity (NR), Gravitational Self-Force (GSF) theory, and, more recently Post-Minkowskian (PM) theory, 
and even Quantum Scattering Amplitudes. The EOB formalism has been the basis of the computation of many of the gravitational wave 
templates \cite{Taracchini:2013rva,Bohe:2016gbl} which have been used in the data analysis of the gravitational wave signals detected by the LIGO and Virgo interferometers \cite{Abbott:2016blz,Abbott:2016nmj,Abbott:2017vtc,Abbott:2017oio}. 

From the practical point of view, the EOB Hamiltonian is made of several building blocks involving various coupling functions or ``potentials."
The EOB potentials were initially computed in PN-expanded form, i.e. as power series 
 in the inverse of the speed of light, $\frac1c$. This PN-expanded approach has, however, several limitations.
 It, indeed,  becomes inaccurate in two regimes that are relevant for theoretically describing present and future gravitational-wave observations, namely: (i)
 when  two bodies of comparable masses become very close to each other, and also, (ii) when a small body moves on a high-energy orbit around
 a large one. The first limitation has been tamed by a combination of analytical resummation methods \cite{Damour:2000we,Damour:2008gu}
 and of strong-field improvements of the EOB potentials by best fitting them to the results of a few NR simulations
 \cite{Buonanno:2007pf,Damour:2007yf,Damour:2007vq}.
The second limitation has been shown to give rise to power-law singularities, at a finite radius corresponding to the light-ring,
 in the small-$\nu$ expansion of the EOB potentials (as they are usually defined) \cite{Akcay:2012ea}.
 
 The latter (light-ring related) limitation was, however, shown to be due to the combined use of a GSF expansion in powers of $\nu$, Eq. \eqref{nu},
 and of a special choice of coordinates in phase-space. By using another way of fixing the phase-space coordinates, and by using the
 exact dependence of the resulting EOB potentials on $\nu$, one can avoid the presence of a singularity at the light ring
 \cite{Akcay:2012ea,Damour:2017zjx}. In order to reach such a conclusion, it was necessary to use a Post-Minkowskian (PM) approach
 to EOB theory \cite{Damour:2016gwp}, i.e. an expansion in powers of the gravitational constant $G$, without using an expansion
 in $\frac1c$ (i.e. without making any slow-motion assumption).
 
 After its introduction in Ref. \cite{Damour:2016gwp}, the PM approach to EOB theory was extended to spin coupling effects
in Ref. \cite{Bini:2017xzy}. The  spin-orbit contribution to the EOB effective Hamiltonian is parametrized by two gyrogravitomagnetic ratios, $g_S$ and $g_{S_*}$: 
\begin{eqnarray}
\label{Heff}
H_{\rm eff}&=&  \sqrt{A\left( \mu^2  +{\mathbf P}^2 +\left(\frac{1}{B}-1\right)P_R^2+Q \right)}\nonumber\\
&+& \frac{G}{ R^3} \left(g_S \, \mathbf{L} \cdot \mathbf{S} + g_{S_*} \, \mathbf{L} \cdot \mathbf{S_*} \right) \,,
\end{eqnarray}
where $\mathbf{S}$ and $ \mathbf{S_*}$ have been defined in Eqs. \eqref{S}, \eqref{Ss}, above, and where  ${\mathbf L}$ denotes the EOB orbital angular momentum 
\beq
{\mathbf L}={\mathbf R}\times {\mathbf P}\,.
\eeq
Here, and below, we use standard vectorial notation for various EOB vectorlike objects.

The values of $g_S$ and $g_{S_*}$ have been computed in Ref. \cite{Bini:2017xzy} at the first post-Minkowskian (1PM) order (i.e., first-order in $G$ but {\it all orders in} $v/c$).  [In view of the factorization of one power of $G$ in the definition of the spin-orbit
contribution to Eq. \eqref{Heff}, this meant computing $g_S$ and $g_{S_*}$ as exact functions of the energy,  at
zeroth order in the gravitational potential $u\equiv GM/(c^2 R)$.] For similar PM results for nonlinear effects in the spins,
see Ref. \cite{Vines:2017hyw}.

As explained in Ref. \cite{Bini:2017xzy}, the PM computation of spin-orbit couplings is achieved by considering the spatial rotations of
the spin vectors of the two gravitationally interacting bodies during the  full scattering process of an hyperboliclike encounter.
The latter spatial rotations are the spin analogs of the orbital scattering angle, and are measured by the \lq\lq spin holonomy" \cite{Bini:2017xzy},
whose definition is recalled below.

The aim of the present paper is to extend the accuracy of the computation of the spin holonomy  to the 
{\it second post-Minkowskian} (2PM) level, and to deduce from it the 2PM-accurate values of the two EOB
gyrogravitomagnetic couplings, i.e.
\begin{eqnarray} \label{gSSsexp1}
g_{S}&=& g_{S}^{\rm 1PM}(H_{\rm eff})+g_{S}^{\rm 2PM}(H_{\rm eff})\, u  +O(u^2)\,,\nonumber\\
g_{S*}&=& g_{S*}^{\rm 1PM}(H_{\rm eff})+g_{S*}^{\rm 2PM}(H_{\rm eff}) \,u +O(u^2)\,,
\end{eqnarray}
where we recall that $u\equiv GM/(c^2 R)= O(G)$. Our 2PM-level computation will make a crucial use of the 2PM-accurate 
explicit analytic computation of the metric generated by two (non-spinning) bodies done long ago by Bel {\it et al.} \cite{Bel:1981be}.

We use a mostly plus spacetime signature, and will often use units where $c=1$. The spin four-vectors of the two bodies are denoted by $s_1$ and $s_2$. The latter four-vectors are (when working, as we do, linearly in spins) orthogonal to the worldlines of the bodies.
The EOB spin three-vectors ${\bf S}_1$, $ {\bf S}_2$ boatin obtained by boosting  $s_1$ and $s_2$ to the center-of-mass frame.
[Viewed in the asymptotically flat spacetime the boosting of the four-vectors $s_1$ and $s_2$ defines corresponding spacetime four-vectors $S_1$ and $S_2$, whose (only non vanishing) spatial components define the EOB spin vectors ${\bf S}_1$ and $ {\bf S}_2$.]

\section{Scattering holonomy}

We consider the gravitational scattering of two massive and spinning bodies {\it at  linear order in spin}. At this order, we can compute the spin holonomy simply from the knowledge of the metric $g(m_1,m_2)=g_{\mu \nu}dx^\mu \otimes dx^\nu$ 
generated by two non spinning masses $m_1$, $m_2$.
In addition, we can neglect the spin-curvature force present in the Mathisson-Papapetrou-Dixon equations, i.e. consider
that each spin vector is parallely propagated in $g_{\mu \nu}(m_1,m_2)$. The explicit value of the metric $g_{\mu \nu}(m_1,m_2)$
generated by the energy-momentum  tensor associated with the two masses $m_1$, $m_2$ has been {\it explicitly} computed
at the 2PM level of approximation in Ref. \cite{Bel:1981be}. [We shall not use here the less explicit 2PM-level 
formulation of the 2PM-accurate metric in Ref. \cite{Westpfahl:1985}.]

The worldline of body 1 (mass $m_1$ and spin $s_1$) is $\mathcal{L}_1$ with $g$-normalized timelike, future-pointing, tangent vector $u_1$ ($u_1\cdot_g u_1\equiv g_{\alpha\beta}u_1^\alpha u_1^\beta=-1$) and linear momentum  $p_1=m_1 u_1$  
($p_1\cdot_g p_1=-m_1^2$). The four-velocity $u_1$ satisfies the geodesic equation
\beq
\frac{D_g u_1}{d\tau^g_1}=0\,,
\eeq 
where $\tau_1^g$ is the proper-time parametrization for $\mathcal{L}_1$ and $D_g$ is the covariant derivative along $u_1$, 
both associated with the metric $g$. 
The spin vector $s_1$ is constrained by the condition of being orthogonal to the linear momentum $p_1$
\beq
\label{s1_dot_p1}
s_1 \cdot_g p_1=0\,,
\eeq
and evolves along $\mathcal{L}_1$ by parallel transport
\beq
\frac{D_g s_1}{d\tau_1^g}=0\,.
\eeq 
As a consequence, the magnitude of $s_1$ is conserved, and the orthogonality  condition \eqref{s1_dot_p1} is preserved.
Corresponding equivalent statements can be done for $u_2$ and $s_2$ and their evolution along $\mathcal{L}_2$. 

Using abstract notation and differential forms, the transport equations can be recast as
\beq 
\label{evolutionsystem}
D_g u_1=0 = D_g s_1 \,.
\eeq
Here
\beq
D_g = d + \omega_1\,,
\eeq
where $d$ denotes the ordinary differential operation along  $\mathcal{L}_1$ while $\omega_1$ denotes the evaluation along $\mathcal{L}_1$ of the
Levi-Civita connection one-form, $\omega$.  
In a coordinate frame the connection one-form acting on contravariant four-vectors is given by
\beq
\omega^\mu_{\; \; \nu}=\Gamma^\mu_{\; \; \nu \lambda}\, dx^\lambda\,,
\eeq
where
\beq
\Gamma^\mu_{\; \; \nu \lambda}= \frac12 g^{\mu \sigma} \left( \partial_\nu g_{\lambda \sigma} + \partial_\lambda g_{\nu \sigma} - \partial_\sigma g_{\nu \lambda} \right)\,.
\eeq

When dealing, as we do here, with the metric generated by {\it point masses}, the metric and the connection are singular
when evaluated on the worldlines. We will use their regularized values, as explicitly discussed and computed in Ref. \cite{Bel:1981be}.

Let us briefly recall the two basic concepts of {\it scattering holonomy} and {\it spin holonomy}, introduced
in Ref. \cite{Bini:2017xzy}. First, the scattering holonomy  along the worldline $\mathcal{L}_1$ is the linear operator   $\Lambda_1$ of 
integrated parallel-transport, acting on contravariant  vectors.  The 
integration along $\mathcal{L}_1$ is performed from the infinite past (i.e., before the interaction, where the spacetime is assumed to be flat) to the infinite future (i.e., after the interaction where again the spacetime is assumed to be flat, since we are studying here an isolated system). 
The two asymptotic Minkowski flat spacetimes are (at this order) unambiguously identified among them, so that $\Lambda_1 = \Lambda_1^\mu{}_{\nu}$ is simply a Lorentz matrix (preserving the Poincar\'e-Minkowski metric $\eta_{\mu\nu}$).  

The solution of the evolution equation of a parallely transported vector $v$ 
\beq
d v= - \omega_1 \, v
\eeq
can be iteratively solved as 
\begin{eqnarray}
v^\mu(\tau)&=&v^\mu(-\infty)-\int_{-\infty}^{\tau} \omega_1^\mu{}_\nu (\tau) v^\nu(\tau)  \nonumber\\
&=&
v^\mu(-\infty)-\int_{-\infty}^{\tau}  \omega_1^\mu{}_\nu (\tau)\left[v^\nu(-\infty) \right. \nonumber\\
&& \left.\qquad -\int_{-\infty}^\tau \omega_1^\nu{}_\sigma (\tau')v^\sigma(\tau')  \right]+\ldots
\end{eqnarray}
so that when $\tau\to+\infty$ we find
\beq
v^\mu_+ =\Lambda_1^\mu{}_{\nu}v^\nu_- \,,
\eeq
where $v^\mu(\pm \infty)=v^\mu_\pm$. The scattering holonomy operator $\Lambda_1$  is therefore given by
\begin{eqnarray} \label{dysonexp}
\Lambda_1  &=& T_{{\mathcal L}_1}[e^{-\int \omega_1}] \nonumber\\
&=& I -\int_{-\infty}^{+\infty} \omega_1  +\frac12 \int_{-\infty}^{+\infty} \int_{-\infty}^{+\infty} T [\omega_1 \omega_1 ']+\ldots\,.
\end{eqnarray}
with $T_{{\mathcal L}_1}$ denoting Dyson's time-ordered product \cite{Dyson:1949bp} along the worldline ${\mathcal L}_1$.
The $T$-ordered integral is such that
\beq \label{dyson2}
\frac12 \iint  T_{{\mathcal L}_1}[\omega_1 \omega_1 ']^\mu{}_\nu= \int_{-\infty}^{+\infty }\omega_1^\mu{}_\sigma(\tau)  \int_{-\infty}^\tau \omega _1{}^\sigma{}_\nu( \tau')  \,.
\eeq
[Note that in our  differential-form
formulation, $\omega_1^\mu{}_\nu(\tau) = \Omega_1^\mu{}_\nu(\tau) d\tau$, the integration does not depend on the choice of any specific parametrization $\tau$ along the worldline.]
The  scattering holonomy map $\Lambda_1$ is then a linear map relating the asymptotic states $u_1^-, s_1^-$ and $u_1^+, s_1^+$:
\beq \label{scatteringoperator}
u_1^+= \Lambda_1 u_1^- \, ;\qquad s_1^+= \Lambda_1 s_1^-  \,.
\eeq

The linear map $ \Lambda_1$ contains information both about the orbital scattering angle, and the spin-rotation angle, of the first particle.
Some extra operation is needed to extract the spin rotation of direct interest for computing the EOB spin-orbit coupling entering the
effective EOB Hamiltonian \eqref{Heff}. What is needed is an operation transforming the
covariant spin four-vectors $s_1$, $s_2$ into objects directly related to the  spatial, canonical  spin three-vectors $\vS_1, \vS_2$ that enter any Hamiltonian dynamics, like the EOB dynamics. As discussed in Ref. \cite{Bini:2017xzy}, $\vS_1$ and $\vS_2$ can
be identified with the four-vectors $S_1$, $S_2$ defined by boosting  $s_1$ and $s_2$, respectively, from their own local rest spaces ($s_1$ is orthogonal to $u_1$ and $s_2$ is orthogonal to $u_2$) onto the local three-space defined by a global field of  (future-directed) unit time-vectors $U = U^\mu e_\mu$ ($e_\mu$ denoting here a coordinate frame) orthogonal to the time-slicing used to describe the full spacetime generated by the two particles.  In a scattering problem, this global field 
 only enters through its asymptotic values at $\pm \infty$, $U=U^{\rm as}$. 
The corresponding asymptotic values  of the  spin vector $S_1$ are then given by
\begin{eqnarray} 
\label{S1as}
S_1^+ &=& B_\eta(u_1^+ \!\!\to \! U^{\rm as}) \, s_1^+ \nonumber\\
S_1^- &=& B_\eta(u_1^- \!\!\to \! U^{\rm as}) \, s_1^- \,,
\end{eqnarray}
with $\eta$  being the Minkowski metric  at $\pm \infty$. Here, the linear map $B_\eta(u \!\!\to \! v)$ is the Lorentz boost matrix
transforming the four-vector $u$ into the four-vector $v$. The explicit expression of the Lorentz matrix 
$\left[B_\eta(u \to v)\right]^\mu{}_\nu$ is (see, e.g., \cite{Bini:2017xzy})
\begin{eqnarray} \label{Buv}
 &&\left[B_\eta(u \to v)\right]^\mu{}_\nu =
\delta^\mu{}_\nu \nonumber \\  \! \! \!  &+& \frac{1}{1-u\cdot v}\left[u^\mu u_\nu +v^\mu v_\nu +  u^\mu v_\nu -  (1-2 u\cdot v) v^\mu u_\nu  \right] . \nonumber \\
\end{eqnarray}
Here the scalar product with respect to the flat metric $\eta$, i.e.,  \lq\lq $\cdot_\eta$," is simply denoted as \lq\lq $\cdot$" to ease the notation.

In the following, we will work, as  is standard in EOB theory, in the center-of-mass (c.m.) frame of the binary system, i.e. we will take for $U^{\rm as}$  the following
unit timelike, asymptotic four-vector
\beq
U^{\rm as}= \left[\frac{p_1+p_2}{{\mathcal E}_{\rm real}}\right]^+= \left[\frac{p_1+p_2}{{\mathcal E}_{\rm real}}\right]^- \,,
\eeq
where, in the conservative dynamics we are considering, the asymptotic values at $\pm \infty$ of the total
four-momenta $(p_1+p_2)^\pm$ coincide. Here, $ {\mathcal E}_{\rm real}$ denotes the total energy
of the binary system (including the rest-mass energy) in the c.m. frame, which is precisely defined as being the (Minkowski) norm of the
asymptotic value of $p_1+p_2$:
\beq \label{s}
s={\mathcal E}_{\rm real}^2 = - (p_1^++ p_2^+)^2 =  - (p_1^-+ p_2^-)^2  \,.
\eeq
In Eq. \eqref{s} we used the traditional notation $s$ for the first Mandelstam variable.

By combining Eqs. \eqref{scatteringoperator}-\eqref{S1as} above, we obtain the linear map between the two asymptotic values (at $\pm \infty$)
of the spatial spin vector of the first particle, namely
\beq
S_1^+= R_1 \, S_1^-  \,,
\eeq
where
\begin{eqnarray} \label{R1}
R_1 &\equiv& B_\eta(u_1^+ \!\!\to \! U^{\rm as}) \, \Lambda_1 \, \left[ B_\eta(u_1^- \!\!\to \! U^{\rm as}) \right]^{-1}\nonumber\\
&=& B_\eta(u_1^+ \!\!\to \! U^{\rm as}) \, \Lambda_1 \,  B_\eta(U^{\rm as} \!\!\to \! u_1^-) \,, 
\end{eqnarray}
with an analogous result for $R_2$.

The linear operator $R_1$ is easily seen to leave $U^{\rm as}$ invariant: $ R_1 U^{\rm as}=U^{\rm as}$. In addition,
as all the linear maps involved in Eq. \eqref{R1} preserve the (Minkowski) length, and as $R_1$ transforms $ S_1^-$
into $S_1^+$ (both spin vectors living in the three-space orthogonal to $U^{\rm as}$),
we conclude that the linear map $R_1$,
which we shall call the {\it spin holonomy} of $\mathcal{L}_1$, is an  $SO(3)$ rotation acting within the three-space orthogonal to $U^{\rm as}$
(in the asymptotic Minkowski space).

We can express the spin holonomy, Eqs. \eqref{R1}, entirely in terms of the {\it incoming} asymptotic values by replacing $u_1^+$ by $ \Lambda_1  u_1^-$,
so that
\beq \label{R1ter}
R_1= B_\eta(\Lambda_1 u_1^- \!\!\to \! U^{\rm as}) \, \Lambda_1 \, \left[ B_\eta(u_1^- \!\!\to \! U^{\rm as}) \right]^{-1}\,.
\eeq

\section{Evaluation of the spin holonomy} \label{sec3}

In order to proceed with the evaluation of the spin holonomy $R_1$, it is convenient to fix a coordinate system as well as an explicit  representation of the asymptotic vectors $u_1$ and $u_2$  at past infinity: $u_1^-$ and $u_2^-$. We assume that the scattering process is confined to the $x-y$ plane of a Lorentzian coordinate system $x^\alpha=(t,x,y,z)$ and that, with respect to this coordinate system, the particle $1$ at past infinity is at rest, while the particle $2$ moves along the negative $y$ direction (so that the angular momentum of the system is aligned with the positive $z$ axis). [In the following, we will generally use greek indices from the beginning of the alphabet
to denote indices pertaining to the special coordinate system $x^\alpha$ attached to the incoming state of particle 1.]
Denoting the corresponding coordinate frame $\partial/\partial x^\alpha$ as $e_\alpha=\{e_t,e_x,e_y,e_z\}$ the two incoming four-velocities read
\beq \label{incomingframe}
u_1^-=e_t \,,\quad u_2^-=\gamma e_t -\sqrt{\gamma^2-1}e_y\,.
\eeq
Here, $\gamma$ denotes the relative Lorentz $\gamma$ factor between the incoming particles, i.e.
\beq
\gamma \equiv - u_1^- \cdot u_2^- \,.
\eeq
It will play an important role in all our computations.

The four-velocity, $U^{\rm as}$, of the c.m. frame then reads
\beq
\label{Uas_def}
U^{\rm as}=\frac{m_1}{{\mathcal E}_{\rm real}}u_1^- +\frac{m_2}{{\mathcal E}_{\rm real}} u_2^-=\cosh \alpha e_t -\sinh\alpha e_y\,,
\eeq
where we have introduced the rapidity parameter $\alpha$ of the boost between the incoming rest frame of particle 1
and the c.m. frame:
\beq
\sinh \alpha = \frac{m_2 \sqrt{\gamma^2-1}}{{\mathcal E}_{\rm real}}\,,\qquad 
\cosh \alpha = \frac{m_1+m_2 \gamma}{{\mathcal E}_{\rm real}}\,.
\eeq
As stated above, in the scattering process $U^{\rm as}=U^{-\rm as}=U^{+\rm as}$ is conserved. Equivalently the total linear four-momentum is conserved, 
\beq
P_{\rm tot}=p_1^-+p_2^-=p_1^++p_2^+\,.
\eeq
Let us also introduce a coordinate frame $e_{\hat \alpha}=\{e_{\hat t},e_{\hat x},e_{\hat y},e_{\hat z}\}$ attached to
the center-of-mass frame, with basis vectors
\begin{eqnarray}
&& e_{\hat t}=U^{\rm as}\,,\quad e_{\hat x}=e_x\,,\nonumber\\
&& e_{\hat y}=-\sinh \alpha e_t +\cosh \alpha e_y\,,\quad e_{\hat z}=e_z\,.
\end{eqnarray}
In this c.m. frame, only the direction of the spatial linear momentum of each particle is changed from $e_{\hat y}^-= e_{\hat y}$ to, say, $e_{\hat y}^+$. 
We can then write
\beq
p_1^-={\mathcal E}_1e_{\hat t} + p\, e_{\hat y}^-\,,\qquad p_1^+={\mathcal E}_1e_{\hat t} + p\, e_{\hat y}^+
\eeq
where $p$ denotes the magnitude of the three-momentum in the c.m. frame (which is, by definition, common to both particles)
\begin{eqnarray}
{\mathcal E}_1&=& \sqrt{m_1^2+p^2}=m_1\cosh \alpha\,,\nonumber\\
  {\mathcal E}_1+{\mathcal E}_2&=& {\mathcal E}_{\rm real}\,,\quad p=m_1 \sinh \alpha\,.
\end{eqnarray}
We recall that ${\mathcal E}_1$ denotes the c.m. energy of the first particle. The change in the direction of the  c.m. three-momentum is measured by the c.m. scattering angle $\chi$. Namely, $\chi$ is the angle of rotation between $e_{\hat y}^-$ and $e_{\hat y}^+$,
so that
\beq
e_{\hat y}^+\cdot e_{\hat y}^-=\cos \chi\,.
\eeq
In addition, $\chi$ enters the second  Mandelstam variable $t = -(p_1^+- p_1^-)^2$ (measuring the invariant momentum transfer),
\beq
\sqrt{-t}=2 p \sin \frac{\chi}{2} \,.
\eeq

Using the definitions above, it is easily checked that
\begin{eqnarray}
u_1^+
&=& \cosh \alpha [\cosh \alpha e_t-\sinh \alpha e_y]\nonumber\\
& +& \sinh \alpha [\sin \chi e_x +\cos  \chi (-\sinh \alpha e_t +\cosh \alpha e_y)]\,,\nonumber\\
\end{eqnarray}
so that the coordinate components $\Lambda_{(1)}^\alpha{}_0$ of $\Lambda_{1}$ in our
special (particle-1-related) coordinate system, which must satisfy
\beq
u_1^+{}^\alpha = \Lambda_{(1)}^\alpha{}_\beta u_1^-{}^\beta = \Lambda_{(1)}^\alpha{}_0\,,
\eeq
are explicitly given by
\begin{eqnarray}
\Lambda_{(1)}^0{}_0&=&  1+\sinh^2 \alpha  (1-\cos \chi )\nonumber\\
\Lambda_{(1)}^1{}_0&=&  \sinh \alpha \sin \chi \nonumber\\
\Lambda_{(1)}^2{}_0&=&  -\sinh \alpha \cosh \alpha  (1-\cos \chi )\nonumber\\
\Lambda_{(1)}^3{}_0&=& 0\,.
\end{eqnarray}

We can then use Eq. \eqref{R1ter}, written in its inverse form, i.e., 
\beq \label{R1inv}
\Lambda_1= \left[B_\eta(\Lambda_1 u_1^- \!\!\to \! U^{\rm as})\right]^{-1} \, R_1 \,  B_\eta(u_1^- \!\!\to \! U^{\rm as}) \,,
\eeq
 to evaluate the remaining components of the matrix $\Lambda_1$. Indeed, we know that the components of
 the rotation matrix  $R_1$ in the center-of-mass frame  (indicated with hatted greek indices) read simply
\beq
R_1{}^{\hat \alpha}{}_{\hat \beta} 
= 
\begin{pmatrix}
1& 0 & 0 &0 \cr
0 & \cos\theta & \sin\theta & 0\cr
0 & -\sin\theta &  \cos  \theta &0 \cr
0 & 0 & 0 & 1\cr
\end{pmatrix}\,.
\eeq
Translating this back to our special coordinate frame $e_\alpha$ one finds
\begin{widetext} 
\beq
R_1{}^\alpha{}_\beta =\begin{pmatrix}
\cos\theta +(\cosh ^2 \alpha-1)(1-\cos\theta)   & \sin \theta \sinh \alpha  & \cosh \alpha \sinh \alpha (1-\cos\theta)  &0 \cr
\sin \theta \sinh \alpha  & \cos \theta  & \sin \theta \cosh \alpha  & 0\cr
-\cosh \alpha \sinh \alpha (1-\cos\theta) & -\sin \theta \cosh \alpha  &   -\cosh ^2 \alpha(1-\cos\theta)+1 &0 \cr
0 & 0 & 0 & 1\cr
\end{pmatrix}\,.  
\eeq
Similarly, using for instance Eq. \eqref{Buv}, one  can determine the coordinate components of the boost matrices $B_\eta(u_1^-\to U^{\rm as})$ and $B_\eta(U^{\rm as}\to u_1^+)$\,, namely
\beq 
B_\eta(u_1^-\to U^{\rm as})^\alpha{}_\beta = \begin{pmatrix}
\cosh \alpha & 0 & -\sinh \alpha &0 \cr
0 & 1 & 0 & 0\cr
-\sinh \alpha & 0 &  \cosh \alpha &0 \cr
0 & 0 & 0 & 1\cr
\end{pmatrix}\,,\qquad
B_\eta(U^{\rm as}\to u_1^-)^\alpha{}_\beta = \begin{pmatrix}
\cosh \alpha & 0 &  \sinh \alpha &0 \cr
0 & 1 & 0 & 0\cr
 \sinh \alpha & 0 &  \cosh \alpha &0 \cr
0 & 0 & 0 & 1\cr
\end{pmatrix}\,,
\eeq
and, 
\begin{eqnarray}
B_\eta(U^{\rm as}\to u_1^+)^\alpha{}_\beta &=&
\begin{pmatrix}
C^2 s^2 (C-1)+(1-s^2) C+s^2& -S s (C c+C-c)& S (s^2 C^2-s^2 C-c)& 0\cr
S s (C c-C-c)& s^2 C+1-s^2& S^2 s \frac{(-1+C c-C)}{(1+C)}& 0\cr
-S (s^2 C^2-s^2 C+c)& S^2 s \frac{(1+C c+C)}{(1+C)}& -C (s^2 C^2-s^2 C-1)& 0\cr
0& 0& 0& 1\cr
\end{pmatrix} \,,
\end{eqnarray}
\end{widetext}
where we shortened the notation as
\beq
[s,c,S,C]=[\sin\chi,\cos\chi,\sinh \alpha,\cosh \alpha]\,.
\eeq 

Using Eq. \eqref{R1inv}, we can then determine the  remaining nonvanishing components of the matrix $\Lambda$ (in our special coordinate frame) as being given by
\begin{eqnarray}
\label{rel_lambdafin}
\Lambda_{(1)}^0{}_0 &=&-c C^2+c+C^2\nonumber\\ 
\Lambda_{(1)}^0{}_1  &=& -S s (-c-C+C c)\cos \theta \nonumber\\
&&+S (-1+c) (C c-c-1)\sin\theta \nonumber\\
\Lambda_{(1)}^0{}_2 &=&  -S (-1+c) (C c-c-1)\cos \theta \nonumber\\
&&-S s (-c-C+C c)\sin \theta  \nonumber\\ 
\Lambda_{(1)}^1{}_0 &=& S s\nonumber\\ 
\end{eqnarray}
\begin{eqnarray}
\Lambda_{(1)}^1{}_1 &=& (-C c^2+C+c^2)\cos \theta \nonumber\\
&&-c s (C-1)\sin \theta  \nonumber\\
 \Lambda_{(1)}^1{}_2 &=& c s (C-1)\cos \theta \nonumber\\
&& +(-C c^2+C+c^2)\sin \theta   \nonumber\\
 \Lambda_{(1)}^2{}_0  &=& C (-1+c) S  \nonumber\\ 
\Lambda_{(1)}^2{}_1  &=& (C-1) s (-C+C c-1)\cos \theta \nonumber\\
&&+(-c-C+c C^2+C c^2-C^2 c^2)\sin \theta  \nonumber\\ 
\Lambda_{(1)}^2{}_2 &=& (C^2 c^2-c C^2-C c^2+c+C)\cos \theta \nonumber\\
&& +(C-1) s (-C+C c-1)\sin \theta  \,.
\end{eqnarray}

The most useful component for our computations below will be
$\Lambda_{(1)}^1{}_2 = \Lambda_{(1)}^x{}_y $. Actually, as we will evaluate this component to the 2PM accuracy only,
we can use a simplified, 2PM-accurate, form of this expression. Indeed, it is easily checked that if one knows  $\chi$ and $\theta$ 
to  their 2PM-expanded accuracy:  $\chi=\chi_{\rm exp}=G\chi_1 +G^2\chi_2 +O(G^3)$ and $\theta=\theta_{\rm exp}=G\theta_1 +G^2\theta_2 + O(G^3)$, we have the sufficiently accurate simplified expression  
\beq \label{lambdavstheta}
 \Lambda_{(1)}^1{}_2  =  \theta_{\rm exp} +(\cosh\alpha-1)\chi_{\rm exp} +O(G^3)\,. 
\eeq
The list of all components in expanded form follows below:
\begin{eqnarray}
\Lambda_{(1)}^0{}_0 &\approx &1+\frac12 (\cosh^2\alpha-1)\chi_1^2 G^2  \nonumber\\ 
\Lambda_{(1)}^0{}_1  &\approx& \sinh \alpha \chi_{\rm exp}\nonumber\\
\Lambda_{(1)}^0{}_2 &\approx& -\sinh\alpha \left[ \left( 1-\frac12 \cosh \alpha\right)\chi_1^2-\theta_1 \chi_1 \right]G^2\nonumber\\ 
\Lambda_{(1)}^1{}_0 &\approx&  \sinh \alpha \chi_{\rm exp} \nonumber\\ 
\Lambda_{(1)}^1{}_1 &\approx& 1+\left[(1-\cosh \alpha)(\chi_1\theta_1-\chi_1^2)-\frac12 \theta_1^2  \right]G^2\nonumber\\
\Lambda_{(1)}^1{}_2 &\approx&   \theta_{\rm exp} -(1-\cosh\alpha)\chi_{\rm exp} \nonumber\\
\Lambda_{(1)}^2{}_0  &\approx& -\frac12 \sinh \alpha \cosh \alpha \chi_1^2 G^2\nonumber\\ 
\Lambda_{(1)}^2{}_1  &\approx&  -\theta_{\rm exp} +(1-\cosh\alpha)\chi_{\rm exp}\nonumber\\ 
\Lambda_{(1)}^2{}_2 &\approx& 1+\left[-\frac12 (1+\cosh^2\alpha)\chi_1^2\right.\nonumber\\
&& \left. +\chi_1\theta_1 (1-\cosh \alpha)-\frac12 \chi_1^2-\frac12 \theta_1^2 \right]G^2 \,.
\end{eqnarray}
As we see, the only components giving 2PM access to the spin rotation angle $\theta=\theta_{\rm exp}$ are $\Lambda_{(1)}^1{}_2$ and $\Lambda_{(1)}^2{}_1$.
 In the next sections we will show how to compute the 2PM-accurate value of  $ \Lambda_{(1)}^1{}_2$ (in our
 special frame). This will allow us to compute the 2PM-accurate value of the spin holonomy rotation angle $\theta$.

\section{The two-body metric at the 2PM order}

As explained above, to compute the spin holonomy one needs to evaluate the connection one-form associated with the metric 
$g_{\mu\nu}(m_1,m_2)$ generated by the two bodies (along the worldline of one of the two bodies, say the body 1).  The 2PM-accurate
value of  $g_{\mu\nu}(m_1,m_2)$ has been explicitly derived (and regularized) long ago in Ref. \cite{Bel:1981be}.
In any given Lorentzian coordinate system $x^\mu$, with associated Minkowski flat metric $\eta_{\mu\nu}={\rm diag}[-1,+1,+1,+1]$,
the solution is conveniently written in terms of the \lq\lq gothic metric" $\mathfrak{g}^{\mu\nu}(x)$ (see Eq. (93) of Ref. \cite{Bel:1981be})
\beq
\label{met_exp}
\mathfrak{g}^{\mu\nu}(x)=\sqrt{-g(x)}g^{\mu\nu}(x)\equiv\eta^{\mu\nu}+ {\sf h}^{\mu\nu}(x)\,.
\eeq
Note that ${\sf h}^{\mu\nu}(x)$ denotes the gothic perturbation of the metric, which differs (but is one-to-one)
related to the usual metric perturbation $h_{\mu\nu}(x)$, defined by writing
\beq
g_{\mu\nu}(x)=\eta_{\mu\nu}+ h_{\mu\nu}(x)\,.
\eeq
The 2PM solution for the metric of the two-body system found in Ref.~ \cite{Bel:1981be}  is explicitly
expressed in terms of  the four-velocities of the bodies, $u_1$ and $u_2$, and their masses. The 
 spacetime dependence of ${\sf h}^{\mu\nu}(x)$  goes through the evaluation of retarded time effects.
 In addition, one should consider only the part of the metric (and the connection) which is \lq\lq regular" in
 the vicinity of the worldline of one of the two bodies. This regular part involves both conservative and radiation-reaction
 effects. We will see that radiation-reaction effects do not contribute to $\theta$, so that our computation based on the
 retarded solution given in Ref.~\cite{Bel:1981be} will give the conservative 2PM value of $\theta$.
 
Following Ref.~\cite{Bel:1981be}, one can write
\begin{eqnarray}
\label{metric2PM}
{\sf h}^{\mu\nu}&=& Gm_1 {\sf h}_{m_1}^{\mu\nu}+Gm_2 {\sf h}_{m_2}^{\mu\nu}\nonumber\\
&+& G^2 m_1^2 {\sf h}_{m_1^2}^{\mu\nu}
+G^2 m_1m_2 {\sf h}_{m_1m_2}^{\mu\nu}+G^2 m_2^2 {\sf h}_{m_2^2}^{\mu\nu}\nonumber\\
&=& \sum_n {\sf h}_{n}^{\mu\nu}\,.
\end{eqnarray}
Each one of these terms was given in \cite{Bel:1981be} by an explicit expression involving either the simply-retarded four-velocities of the two bodies, or  some doubly-retarded quantities (involving two successive retarded propagations), or (for the most nonlinear terms) the explicit result
of a complicated integration involving the cubic vertex of Einstein's action (``$P$-terms").
We follow the notation and conventions of  Ref.~\cite{Bel:1981be}, apart from the fact that,  to follow our previous 1PM work,
we shall often use labels 1 and 2 to respectively denote the two masses, by contrast to the
unprimed and primed notations used  in Ref.~\cite{Bel:1981be}, where the masses were denoted $m$ and $m'$,
and the corresponding four-velocities $u$ and $u'$, etc.  

The main steps necessary to perform the spin holonomy computation are listed below. We recall that our computations
will finally be done in a special coordinate frame, see Eq. \eqref{incomingframe}, linked to the incoming state of the first particle.
We will evaluate the specific component $\Lambda_{(1)}^1{}_2 = \Lambda_{(1)}^x{}_y $ of the scattering holonomy
in that special frame.

\subsection{ 1PM-accurate orbits of the two bodies}

The solution for $u_1(\tau_1)$  and $u_2(\tau_2)$ parametrized by their (Minkowski) proper times $\tau_1$ and $\tau_2$ is obtained by solving the geodesic equations of the 2PM metric. [As it is customary in PM computations, we use flat space normalization of vectors, \lq\lq$\cdot_\eta$",  and flat space proper time definitions.]
The 2PM equations of motion are explicitly given in Ref.~\cite{Bel:1981be} (see Eqs. (111) there).
Here, we will only need  their solution at the 1PM order of accuracy. Obtaining this solution is straightforward and leads (in
an arbitrary frame) to
\begin{eqnarray}
u_1^\mu(\tau_1)  &=&  u_1^-{}^\mu +G m_2  \frac{(1-2\gamma^2)S(\tau_1)}{\sqrt{\gamma^2-1}D(\tau_1)}\, e_x^\mu    \nonumber\\
&& -Gm_2 \frac{\gamma (2\gamma^2-3)}{\sqrt{\gamma^2-1}D(\tau_1)}  e_y^\mu\,,
\end{eqnarray}
and  
\begin{eqnarray}
z^\mu(\tau_1)-z_1^\mu(0) 
&=& u_1^-{}^\mu \tau_1  +Gm_2 (1-2\gamma^2) \frac{(S(\tau_1)-1)}{(\gamma^2-1)}\, e_x^\mu \nonumber\\
&-& 
Gm_2\frac{\gamma (2\gamma^2-3) }{ \gamma^2-1 }  
\ln \left( S(\tau_1)\right)\,  e_y^\mu\,.
\end{eqnarray}
Here we have introduced the functions $D(\tau)$ and $S(\tau)$ defined by
\beq
D(\tau) \equiv \sqrt{b_0^2 +\tau^2 (\gamma^2-1)}\,,\qquad D(0)=b_0\,,
\eeq
and
\beq
S(\tau)\equiv \frac{1}{b_0}\left(\tau \sqrt{\gamma^2-1} +D(\tau)\right)\,,\qquad S(0)=1\,.
\eeq
Note  that $S(\tau)S(-\tau)=1$.

Similarly,  the unit tangent vector to the worldline of body 2 is given by
\begin{eqnarray}
u_2{}^\mu(\tau_2)&=& u_2^- {}^\mu   -G m_1 \frac{(1-2\gamma^2)S(\tau_2)}{\sqrt{\gamma^2-1}D(\tau_2)}\,  e^\mu_x  \nonumber\\  
&+& Gm_1  \frac{\gamma(2\gamma^2-3)}{(\gamma^2-1)D(\tau_2)}  v'{}^\mu\,, 
\end{eqnarray}
and the worldline  
\begin{eqnarray}
z_2{}^\mu(\tau_2)-z_2{}^\mu(0) &=& u_2^-{}^\mu \tau_2  -G m_1 (1-2\gamma^2)\frac{(S(\tau_2)-1)}{(\gamma^2-1)}\, e^\mu_x  \nonumber\\
&+& Gm_1 \frac{\gamma (2\gamma^2-3)}{(\gamma^2-1)^{3/2}}  
\ln \left(S(\tau_2) \right)\, v'{}^\mu  \,.
\end{eqnarray}
Here  the four-vector $v'$ is defined as (where $P(u)=\eta + u \otimes u $ denotes the projector orthogonal to $u$)
\beq
v' \equiv P(u_2^-)u_1^-=-(\gamma^2-1)e_t +\gamma \sqrt{\gamma^2-1}e_y\,.
\eeq
It is the $1 \leftrightarrow 2$ analog of the four-vector $v \equiv  P(u_1^-)u_2^-= - \sqrt{\gamma^2-1}e_y $.
In the solutions above, we have fixed initial conditions at $\tau=0$ so that
\beq
z_1(0)=b_0 e_x \,,\qquad z_2(0)=0\,.
\eeq
Note that the intrinsic definition of the vector $b_0 e_x$ is such that $z_1(\tau_1)- z_2(\tau_2) = b_0 e_x $
when $\tau_1=\tau_2=0$, which corresponds to connecting the two 1PM-accurate worldlines by a bi-podal line orthogonal
to both worldlines.
The parameter $b_0$ (which is a relativistically defined 1PM-accurate closest distance of approach)
differs by an $O(G)$ term from the usually defined c.m. impact parameter $b$ (defined by the condition that the c.m.
 angular momentum $L$ is equal to $L= b \,p$, where $p$ is the c.m. three-momentum). We derive in
 Appendix D the following relation between $b$ and $b_0$:
\beq \label{bvsb0}
b=b_0 +G(m+m')\frac{(2\gamma^2-1)}{\gamma^2-1} + O(G^2)\,.
\eeq
This relation will be crucial for relating our computation to EOB theory.

Having in hands such explicit expressions for the (1PM-accurate) worldlines allows one to explicitly compute the needed retarded times,
and associated retarded quantities, entering the 2PM metric.
Indeed, given a generic spacetime point $x$ one can first define  the intersection of the past light cone from $x$ with the first worldline $\mathcal{L}_1$.
This defines  a point $z_1^R(x)$ on $\mathcal{L}_1$, corresponding to a value $\tau_1^R(x)$ of the proper time. The null condition 
\beq
(x-z_1(\tau_1^R))^2=0\,,
\eeq
defines the functional link $\tau_1^R=\tau_1^R(x)$ etc. Then, from the point $z_1(\tau_1^R)$ on $\mathcal{L}_1$ one can  draw a second 
past light cone which intersects the worldline $\mathcal{L}_2$ of  body 2 at the doubly retarded point
$z_2^{RR}=z_2(\tau_2^{RR}(x))$. One can determine this doubly-retarded point on $\mathcal{L}_2$ by using the null condition
\beq
(z_1(\tau_1^R)-z_2(\tau_2^{RR}))^2=0\,,
\eeq
thereby obtaining $\tau_2^{RR}$ as a function of $\tau_1^R(x)$ and hence as a function of the spacetime point $x$  itself.

\subsection{Regular terms in the 2PM metric, and its derivative}

Let us henceforth consider the explicit evaluation of 2PM metric (and its derivative) in the special frame where we shall perform our
computation of the scattering holonomy $\Lambda_1$ (hence our use of greek indices from
the beginning of the alphabet). We recall that, in order to compute $\Lambda_1$,
 we must evaluate the regularized value of the metric, and of the Levi-Civita connection, along the first worldline $\mathcal{L}_1$. 
 For detailed discussions of regularization within a PM framework see, e.g., Refs \cite{Damour:1975uj} and \cite{Bel:1981be}.

At first order in $G$, we have the following contributions to the 2PM metric \eqref{metric2PM}:
\begin{eqnarray}
\label{metric2PM}
Gm_1 {\sf h}_{m_1}^{\alpha\beta}&=& -4G m_1 \left(\frac{u_1^\alpha u_1^\beta}{r_1}\right)_R\nonumber\\
Gm_2 {\sf h}_{m_2}^{\alpha\beta}&=&  -4G m_2 \left(\frac{u_2^\alpha u_2^\beta}{r_2}\right)_R\,,
\end{eqnarray}
where
\begin{eqnarray}
r_1 &=& -(x-z_1(\tau_1))\cdot u_1(\tau_1)\nonumber \\
r_2 &=& -(x-z_2(\tau_2))\cdot u_2(\tau_2)\,,
\end{eqnarray}
with the label $R$ denoting the evaluation at the retarded time $\tau_1^R(x)$ etc.
Going to the worldline of  body 1 the term $Gm_1 {\sf h}_{m_1}^{\alpha\beta}$ is singular.  However, $Gm_1 {\sf h}_{m_1}^{\alpha\beta}$ also contains a regular part. Using, e.g.,
Ref. \cite{Damour:1975uj} for the evaluation of this regular part, and of its derivatives, one can explicitly check that 
the regular part of $Gm_1 {\sf h}_{m_1}^{\alpha\beta}$ does not contribute to our computation.
The second contribution, $Gm_2 {\sf h}_{m_2}^{\alpha\beta}$ is purely regular on the worldline $\mathcal{L}_1$.
We must keep in mind that, in view of the 1PM-accurate definition of the retarded propertime $\tau_2(x)$,
it will contribute (when PM-expanded) both at order $G$ and at order $G^2$. 

At the second order in $G$, we have both \lq\lq square" terms
\begin{eqnarray}
\label{metric2PM}
G^2 m_1^2 {\sf h}_{m_1^2}^{\alpha\beta}&=& -G^2 m_1^2 \left(\frac{7u_1^\alpha u_1^\beta +n_1^\alpha n_1^\beta}{r_1^2}\right)_R\nonumber\\
G^2 m_2^2 {\sf h}_{m_2^2}^{\alpha\beta}&=& -G^2 m_2^2 \left(\frac{7u_2^\alpha u_2^\beta +n_2^\alpha n_2^\beta}{r_2^2}\right)_R\,,
\end{eqnarray}
where
\begin{eqnarray}
n_1{}_R (x) &=& \frac{P(u_1{}_R)(x-z_1^R)}{r_1{}_R} \nonumber\\
n_2{}_R (x) &=& \frac{P(u_2{}_R)(x-z_2^R)}{r_2{}_R} \,,
\end{eqnarray}
and \lq\lq mixed" terms proportional to  $G^2 m_1m_2 {\sf h}_{m_1m_2}^{\alpha\beta}$.
We again find that the singular contribution  $G^2 m_1^2 {\sf h}_{m_1^2}^{\alpha\beta}$ 
does not contribute to our calculation.

Finally among the  \lq\lq  mixed" terms we distinguish three different contributions:
\begin{eqnarray}
G^2 m_1m_2 {\sf h}_{m_1m_2}{}^{\alpha\beta}&=& G^2 m_1m_2 {\sf h}^{1/(r\rho)}_{m_1m_2}{}^{\alpha\beta}\nonumber\\
&&+G^2 m_1m_2 {\sf h}^{1/(r'\rho ')}_{m_1m_2}{}^{\alpha\beta}\nonumber\\
&&+G^2 m_1m_2 {\sf h}^P_{m_1m_2}{}^{\alpha\beta}\,,
\end{eqnarray} 
with
\begin{eqnarray}
G^2 m_1m_2 {\sf h}^{1/(r\rho)}_{m_1m_2}{}^{\alpha\beta}&=& -4 G^2 m_1 m_2(1+2\gamma^2)\left( \frac{u_1^\alpha u_1^\beta}{r_1\rho_1} \right)_R\nonumber\\
G^2 m_1m_2 {\sf h}^{1/(r'\rho ')}_{m_1m_2}{}^{\alpha\beta}&=&-4 G^2 m_1 m_2(1+2\gamma^2)\left( \frac{u_2^\alpha u_2^\beta}{r_2\rho_2} \right)_R\nonumber\\
\end{eqnarray}
where
\begin{eqnarray}
\rho_1{}_R &=& -(z_1^R-z_2^{RR})\cdot u_2{}_R \nonumber\\
\rho_2{}_R &=& -(z_2^R-z_1^{RR})\cdot u_1{}_R \,.
\end{eqnarray}
Again, of these two terms the first one (involving $r_1$) is singular and can be discarded in our calculation.
Finally the last ``P"-term, $G^2 m_1m_2 {\sf h}^P_{m_1m_2}{}^{\alpha\beta}$ can be written as
\begin{eqnarray}
G^2 m_1m_2 {\sf h}^P_{m_1m_2}{}^{\alpha\beta}&=&G^2 m_1m_2 [M_1^{\alpha\beta}{}_{\rho\sigma}D^\rho P^\sigma\nonumber\\
&& +M_2^{\alpha\beta}{}_{\rho\sigma}D'{}^\rho P'{}^\sigma\nonumber\\
&& +N_1^{\alpha\beta}{}_{\rho\sigma}D'{}^\sigma P^\rho\nonumber\\
&& +N_2^{\alpha\beta}{}_{\rho\sigma}D^\sigma P'{}^\rho
]_{R }\,,
\end{eqnarray}
where
\begin{eqnarray}
M_1^{\alpha\beta}{}_{\rho\sigma} &=& - 16  u_1^\alpha u_1^\beta u_2{}_\rho u_2{}_\sigma  \nonumber\\
N_1^{\alpha\beta}{}_{\rho\sigma}&=&  8(2u_1^\alpha u_2{}^\beta-\gamma \eta^{\alpha\beta})u_2{}_\rho u_1{}_\sigma\nonumber\\
&& +[-16\gamma u_1^\alpha u_2{}^\beta-2\eta^{\alpha\beta}(2\gamma^2-1)]\eta_{\rho\sigma}\nonumber\\
&&+  16 \gamma u_2{}^\alpha u_1{}_\sigma \delta^\beta_\rho +16 \gamma u_2{}^\beta u_1{}_\sigma \delta ^\alpha_\rho\nonumber\\
&& +4(2\gamma^2-1)\delta^\alpha_\rho \delta^\beta_\sigma \,,
\end{eqnarray}
and where the derivative $D$ and $D'$ are \lq\lq worldline" derivatives, defined in Eq. (29) of Ref.~\cite{Bel:1981be}.
Essentially, $h^\mu D_\mu$ (acting on some functional of the two worldlines, and a function of some given spacetime point $x$)
denotes the geometric operation consisting in infinitesimally translating (as a whole) the
first worldline $\mathcal{L}_1$ by the vectorial amount $h^\mu$. In view of the overall translation invariance of our problem,
the combined derivative $\partial_\mu + D_\mu + D'_\mu$ yields a vanishing result on any functional entering our problem.
We note also that when $ D_\mu$ acts on a functional defined (with sufficient accuracy) from straight worldlines (as
is the case for $P_\mu$ and $P'_\mu$),  the longitudinal derivatives $u^\mu D_\mu$ and $u'^\mu D'_\mu$ yield vanishing
results.

The four-vectors $P_\mu$ and $P'_\mu$ entering the definition of ${\sf h}^P_{m_1m_2}{}^{\alpha\beta}$ (which were one
of the crucial new results of Ref.~\cite{Bel:1981be}) are defined by nontrivial integrals (involving the cubically nonlinear
gravitational vertex), and were explicitly computed in Ref.~\cite{Bel:1981be} (see Appendix C there). In addition, Appendix C there
also gave all needed explicit formulas for evaluating the regularized values of  $P_\mu$ and $P'_\mu$, and their first
derivatives, on the worldlines. See Eqs. (C21)-(C42) there.

As an example of the explicit expressions derived from the above results, we list below the nonvanishing components of the  
\lq\lq P" part of the  metric with respect to our special chosen coordinate system:
\begin{widetext}
\begin{eqnarray}
{\sf h}^P_{m_1m_2}{}^{tt} &=& -16(\gamma^2-1) D^2 P^2+4(3-10\gamma^2)D^{'2}P^2-4(1+6\gamma^2)D^{'1}P^1-16(\gamma^2-1)D^{'2}P^{'2}\nonumber\\
{\sf h}^P_{m_1m_2}{}^{tx} &=&4\frac{(2\gamma^2+1)\sqrt{\gamma^2-1}}{\gamma} D^{'2}P^1-16\gamma \sqrt{\gamma^2-1}D^{'1}P^2 \nonumber\\
{\sf h}^P_{m_1m_2}{}^{ty} &=& 16\gamma \sqrt{\gamma^2-1}D^{'1}P^1+12\frac{\sqrt{\gamma^2-1}(2\gamma^2-1)}{\gamma}D^{'2}P^2+16\frac{(\gamma^2-1)^{3/2}}{\gamma}D^{'2}P^{'2}\nonumber\\
{\sf h}^P_{m_1m_2}{}^{xx} &=& 4(2\gamma^2-1)D^{'1}P^1+4(2\gamma^2-3)D^{'2}P^2\nonumber\\
{\sf h}^P_{m_1m_2}{}^{xy} &=& 4(2\gamma^2-1) D^{'1}P^2 -4(2\gamma^2-3)D^{'2}P^{1} \nonumber\\
{\sf h}^P_{m_1m_2}{}^{yy} &=& -4(2\gamma^2-1)D^{'1}P^1+4(3-2\gamma^2)D^{'2}P^2-16\frac{(\gamma^2-1)^2}{\gamma^2}D^{'2}P^{'2}\nonumber\\
{\sf h}^P_{m_1m_2}{}^{zz} &=& 4(-2\gamma^2+1)D^{'1}P^1+4(2\gamma^2-3)D^{'2}P^2\,.
\end{eqnarray}
\end{widetext}

\section{Computation of $\Lambda_{(1)}^x{}_y =\Lambda_{(1)}^1{}_2$} \label{sec5}

Having discussed the evaluation of the various regular terms entering the 2PM metric, we can now
compute the connection one-form and, thereby (after integration) the 
specific scattering  holonomy component $\Lambda_{(1)}^x{}_y =\Lambda_{(1)}^1{}_2$
necessary to obtain the rotation angle of the center-of-mass spin vector after the full scattering process.

The specific component $\Lambda_{(1)}^1{}_2$ we wish to compute can be decomposed as the sum of various contributions.
On the one hand, it contains the contributions obtained by inserting  the different terms entering the 2PM metric, Eq. \eqref{metric2PM}, 
into the linear piece, $-\int_{-\infty}^\infty \omega_1$, of the expanded  time-ordered exponential \eqref{dysonexp}, say
\begin{eqnarray}\label{dysonlin}
\Lambda_{(1 \,{\rm lin})}^1{}_2&=& \sum_n \Lambda_{(1)}({\sf h}_n)^1{}_2\nonumber\\
&\equiv &
\sum_n \int_{-\infty}^{+\infty} \Gamma({\sf h}_n)^1{}_{2\alpha} u_1^\alpha  d\tau \,.
\end{eqnarray}
On the other hand, one must also add (at the order $G^2$ at which we are working) the quadratic contribution in the Dyson expansion
\eqref{dysonexp}, say 
\beq \label{dysonquad}
\Lambda_{(1 \,{\rm quad})}^1{}_2= \int_{-\infty}^\infty d\tau\omega_{(1)}^1{}_\sigma(\tau)  \int_{-\infty}^\tau d\tau'\omega _{(1)}{}^\sigma{}_2( \tau')  \,.
\eeq

All the terms in Eq. \eqref{dysonlin} must be $G$-expanded, up to the order $O(G^2)$ included (and regularized, as explained above). The 
Christoffel symbols written directly in terms of the gothic metric are given in Eqs. (A12)-(A14) of Ref.~\cite{Bel:1981be}. 
We can distinguish four regularized contributions to $\Lambda_{(1)}^1{}_2$: ${\sf h}_1=Gm_2 {\sf h}_{m_2}^{\alpha\beta}$
contributes to both first and second order in $G$; while one gets only   $O(G^2)$ contributions from
${\sf h}_2=G^2 m_2^2 {\sf h}^{1/(r_2{}^2)}_{m_2^2}{}^{\alpha\beta}$, 
${\sf h}_3=G^2 m_1m_2 {\sf h}^{1/(r_2\rho_2)}_{m_1m_2}{}^{\alpha\beta}$ and
${\sf h}_4=G^2 m_1m_2 {\sf h}^P_{m_1m_2}{}^{\alpha\beta}$.
Explicitly, using the notation
\beq
u_1=u_1^-+\delta u_1\,, 
\eeq
with (in our special coordinate system) $u_1^-{}^\alpha=\delta^\alpha_0$ and $\delta u_1^\alpha=\delta u_1^x \delta^\alpha_x+\delta u_1^y \delta^\alpha_y$, 
we find 
\begin{eqnarray}
\Gamma({\sf h})^1{}_{2\alpha} u_1^\alpha &=& \Gamma({\sf h}_1)^1{}_{2\alpha} (u_1^-{}^\alpha + \delta u_1^\alpha)\nonumber\\
&& +\sum_{n=2}^4 \Gamma({\sf h}_n)^1{}_{2\alpha} u_1^-{}^\alpha\,,
\end{eqnarray}
with
\begin{eqnarray}
\label{Gamma_h1_all}
\Gamma({\sf h}_1)^1{}_{2\alpha} u_1^\alpha&=&  -\frac12 \partial_x {\sf h}_1^{ty}\nonumber\\
&& +\frac14 \left(\partial_y {\sf h}_1^{yy}-\partial_y {\sf h}_1^{tt}  \right) \delta u_1^x \nonumber\\
&& +\frac14 \left(\partial_x {\sf h}_1^{tt}+\partial_x {\sf h}_1^{yy}  \right) \delta u_1^y \nonumber\\
&& +\frac12 \left({\sf h}_1^{yy}-{\sf h}_1^{tt}  \right)\partial_x {\sf h}_1^{ty} \nonumber\\
&&  -\frac14 {\sf h}_1^{ty} \left(\partial_x {\sf h}_1^{tt}-\partial_x {\sf h}_1^{yy} \right)\,,
\end{eqnarray}
and
\begin{eqnarray}
\sum_{n=2}^4 \Gamma({\sf h}_n)^1{}_{2\alpha} u_1^\alpha&=&  -\frac12 \sum_{n=2}^4  \left(\partial_x  {\sf h}_n^{ty}-\partial_y  {\sf h}_n^{tx}\right)\,.
\end{eqnarray}
For example, the second order contributions from ${\sf h}_{2,3,4}$   arise from
\beq
\label{h234}
\int_{-\infty}^\infty \Gamma({\sf h}_n)^1{}_{2\alpha} u_1^-{}^\alpha d\tau_1  = -\frac12 \int_{-\infty}^\infty dt \left( \partial_x {\sf h}_n^{ty} -\partial_y  {\sf h}_n^{tx} \right)\,.
\eeq
They can be computed relatively easily since, having a $O(G^2)$ integrand, they can be integrated along the 0PM approximation of the worldline of  body 1, i.e. a straight worldline: $u_1=u_1^-$, with $u_1^-{}^\alpha=\delta^\alpha_0$ with the choice of coordinates adopted here. We then find
\begin{eqnarray}
\int_{-\infty}^\infty \Gamma({\sf h}_2)^1{}_{2\alpha} u_1^-{}^\alpha d\tau_1  &=& \frac{15}{4}\pi G^2 m_2^2 \frac{\gamma}{b_0^2}\nonumber\\
\int_{-\infty}^\infty \Gamma({\sf h}_3)^1{}_{2\alpha} u_1^-{}^\alpha d\tau_1  &=&   \pi G^2 m_1 m_2\frac{\gamma}{b_0^2}  (1+2\gamma^2) \nonumber\\
\int_{-\infty}^\infty \Gamma({\sf h}_4)^1{}_{2\alpha} u_1^-{}^\alpha d\tau_1  &=& -3  \pi G^2 m_1 m_2  \frac{\gamma}{b_0^2}\,.
\end{eqnarray}
Particular care should be used, however, for the  term ${\sf h}_4=G^2 m_1m_2 {\sf h}^P_{m_1m_2}{}^{\alpha\beta}$. One convenient
way to proceed is the following. One only needs  the two metric components
\beq
\label{rel_h01}
{\sf h}^P_{m_1m_2}{}^{tx}=  -4(1+2\gamma^2 )D'{}^0P^1 -16 \gamma \sqrt{\gamma^2-1} D'{}^1P^2 
\eeq
\begin{eqnarray}
\label{rel_h02}
{\sf h}^P_{m_1m_2}{}^{ty}
&=& -12(2\gamma^2-1)D'{}^0P^2\nonumber\\
&& + 16 \gamma \sqrt{\gamma^2-1}(D'{}^0P'{}^0+D'{}^2 P^2) \nonumber\\
&& + 16 \gamma \sqrt{\gamma^2-1} D'_\alpha P^\alpha\nonumber\\
&\equiv& \tilde {\sf h}^P_{m_1m_2}{}^{ty}+ 16 \gamma \sqrt{\gamma^2-1} D'_\alpha P^\alpha\,,
\end{eqnarray}
to be used in 
\beq
\Gamma({\sf h}_4)^1{}_{2\alpha} u_1^-{}^\alpha=  -\frac12 \left(\partial_x  {\sf h}_4^{ty}-\partial_y  {\sf h}_4^{tx}\right)\,.
\eeq
The regular part of the last term in ${\sf h}^P_{m_1m_2}{}^{ty}$ can be rewritten as \cite{Bel:1981be}
\beq
\left[D'_\alpha P^\alpha\right]_{\rm reg}=-\frac{1}{2r_2\rho_2}\,.
\eeq
Its contribution to $\Lambda_{(1)}{}^1{}_2$ is of the same type as that of ${\sf h}_3$ and can be evaluated in the same way,  
leading to the contribution
\beq
-2 G^2 m_1 m_2 \pi \frac{\gamma}{b_0^2}\,.
\eeq
The contribution due to ${\sf h}^P_{m_1m_2}{}^{tx}$  and the remaining term, $\tilde {\sf h}^P_{m_1m_2}{}^{ty}$, 
in $ {\sf h}^P_{m_1m_2}{}^{ty}$,
\beq
-\frac12 \left(\partial_x  \tilde {\sf h}_4^{ty}-\partial_y  {\sf h}_4^{tx}\right)
\eeq
 leads to the additional term
\beq
\label{additional}
-G^2 m_1 m_2 \pi \frac{\gamma}{b_0^2}\,.
\eeq
This last computation is performed by: 1) exchanging the partial derivatives with the worldline derivatives; 2) replacing $P$ and $P'$ with 
their regular parts using  Eqs. (C21) and (C38) of Ref. \cite{Bel:1981be},
\begin{eqnarray}
&& \partial_x P_x =R^{(1)}{}_{xx}\,,\quad
\partial_y P_x =R^{(1)}{}_{xy}\,,\nonumber\\
&& \partial_x P_y =R^{(1)}{}_{yx}\,,\quad
\partial_y P_y =R^{(1)}{}_{yy}\,,
\end{eqnarray}
and
\beq
\partial_x P'_0 =R'{}^{(1)}{}_{x}\,,
\eeq
taking into account that $R^{(1)}{}_{xy}=R^{(1)}{}_{yx}$ and
\begin{eqnarray}
R^{(1)}{}_{xy}&=& -\frac{1}{4\rho^2}\left[-\frac{A_xv_y}{A}+\left(A+\frac{A_y v_y}{A}\right)\frac{A_x A_y}{A^2}\right]\nonumber\\
R^{(1)}{}_{xx}&=& -\frac{1}{4\rho^2}\left[ \frac{A_x^2}{A}+\frac{A_x^2A_y v_y}{A^3}-A+\frac{A_yv_y}{A} \right]\nonumber\\
R^{(1)}{}_{yy}&=& -\frac{1}{4\rho^2}\left[ -\frac{A_yv_y}{A}+\frac{A_y^2}{A}+\frac{A_y^3 v_y}{A^3}-A \right]\nonumber\\
R'{}^{(1)}{}_{x}&=&\frac{1}{2\rho^2}\left[ (B_x-A_x)\ln A\right.\nonumber\\
&&\left. +\frac12 A_x \left( 1+\frac{2B}{A}-\frac{2}{A^2}\right)
 +\frac{B_x}{2}  \right]\,.
\end{eqnarray}
Here, entered the components of the following two $u_1$-orthogonal vectors
\beq
{\mathbf A}=P(u_1)(u_2+\nu)\,,\qquad {\mathbf B}=P(u_1)(u_2-\nu)
\eeq
as well as
\begin{eqnarray}
\nu &=& \frac{P(u_2)[z_{1R}-z_{2RR}]}{\rho} \,,\nonumber\\
\rho &=& -[z_{1R}-z_{2RR}]\cdot u_2\,.
\end{eqnarray}
The above quantities then become functions of $\nu^x$, $\nu^y$ and $\rho$ and one can use the expressions for the worldline derivatives of the components of $\nu$
and $\rho$, given in Eqs. (C34) and (C36) of Ref. \cite{Bel:1981be}. The result  \eqref{additional} then follows  straightforwardly.

Besides the subtle (and intricate) $P$-contribution discussed above, another delicate evaluation concerns the contribution due to 
the (regularized) linearized gothic metric ${\sf h}_1$,  namely
\begin{eqnarray} \label{h1original}
{\sf h}_1&=& Gm_2 {\sf h}_{m_2}^{\alpha\beta}= -4G m_2 \left(\frac{u_2^\alpha u_2^\beta}{r_2}\right)_R\nonumber\\
&=& -4G m_2 \frac{u_2(\tau_{2R}(x)) ^\alpha u_2(\tau_{2R}(x))^\beta}{r_2( \tau_{2R}(x)) } \,.
\end{eqnarray}
Indeed, this evaluation must be treated with 2PM accuracy, which means taking into account several $O(G)$ fractional 
modifications beyond the pure 1PM evaluation performed in Ref. \cite{Bini:2017xzy}.

The 2PM-accurate contribution associated with ${\sf h}_1$ has several origins. There is a piece coming from
the fact that the first worldline $\mathcal{L}_1$ is curved at the $O(G)$ level. The corresponding contribution,
proportional to $\delta u_1$, was already explicitly written out in Eq. \eqref{Gamma_h1_all}. The last two lines in
Eq. \eqref{Gamma_h1_all} exhibit an ${\sf h}_1$-related contribution that is quadratic in ${\sf h}_1$.
In addition, one should take into account 
$O(G^2)$ corrections to the pure 1PM evaluation of  $\Lambda_{(1)}^1{}_2({\sf h}_1)$
coming from the various $O(G)$ fractional corrections to the retardation effects in $u_2$ and $r_2$,
which enter the definition of ${\sf h}_1$. Indeed, one can formally decompose $\tau_{2R}(x)$ into the sum of
an $O(G^0)$ piece (evaluated as if $\mathcal{L}_2$ was straight) and an $O(G^1)$ one (taking into account
the $O(G)$ curvature of $\mathcal{L}_2$), say
\beq \label{tau2Rexp}
\tau_{2R}(x)=\tau_{2R}^{(0)}(x)+G \delta \tau_{2R}^{(1)}(x) + O(G^2)\,.
\eeq
We find
\begin{eqnarray}
\tau_{2R}^{(0)}(x) &=& \gamma t +\sqrt{\gamma^2-1} y \nonumber\\
&&-\sqrt{[\sqrt{\gamma^2-1}t +\gamma y]^2+x^2}\,,
\end{eqnarray}
where the first-order correction $\delta \tau_{2R}^{(1)}(x)$  can be straightforwardly computed from the $O(G)$ term in 
the $G$-expansion of the defining equation $(x-z_2(\tau_{2R}))^2=0$.
As a consequence of the expansion \eqref{tau2Rexp}, we have corresponding $G^0+ G^1$ expansions for
\beq
u_2(\tau_{2R}(x)) ^\alpha=u_2^-{}^\alpha +G \delta u_2^\alpha (x) + O(G^2)\,,
\eeq
and
\beq
r_2( \tau_{2R}(x)) =\sqrt{[\sqrt{\gamma^2-1}t +\gamma y]^2+x^2} +G \delta r_2(x) + O(G^2)\,.
\eeq
Inserting these expansions in powers of $G$ in the original expression of ${\sf h}_1$ Eq. \eqref{h1original},
yields the following decomposition of ${\sf h}_1$:
\beq \label{h1exp}
{\sf h}_1=G {\sf h}_1^0 + G^2 {\sf h}_1^{\delta u_2}+G^2 {\sf h}_1^{\delta r_2}\,.
\eeq
The term $G {\sf h}_1^0 $ in the latter equation is the linearized metric generated by a straight worldline $\mathcal{L}_2$.
When inserted in the first term in Eq. \eqref{Gamma_h1_all}, namely
\beq
 -\frac12 \partial_x {\sf h}_1^{ty}\,,
\eeq 
and integrated, it yields the pure 1PM contribution to $\Lambda_1$, namely
\beq
{\Lambda^{1 \rm PM}_{(1)}}^1{}_2= 4 G m_2 \frac{ \gamma}{b_0}\,.   
\eeq
Several $O(G^2)$ corrections to this result follow from Eq. \eqref{Gamma_h1_all}.
First, the terms proportional to $\delta u_1$ in Eq. \eqref{Gamma_h1_all} yield
\begin{eqnarray}
&&\int_{-\infty}^\infty \Gamma({\sf h}^{0,\rm lin}_1)^1{}_{2\alpha} \delta u_1^\alpha d\tau_1  =  \nonumber\\
&& \qquad+ G^2 m_2^2 \frac{\gamma}{b_0^2}\left[\pi (1+2\gamma^2)-4 \frac{(2\gamma^2-1)}{\gamma^2-1} \right].
\end{eqnarray}
We must then add the contribution to $\Lambda_{(1)}^1{}_2$ coming from the terms 
$G^2 {\sf h}_1^{\delta u_2}+G^2 {\sf h}_1^{\delta r_2}$ in Eq. \eqref{h1exp}. These terms
can be computed as the ones in Eqs. \eqref{h234}, and yield
\begin{eqnarray}
&&\int_{-\infty}^\infty \Gamma({\sf h}_1^{\delta u_2})^1{}_{2\alpha} u_1^-{}^\alpha d\tau_1  = \nonumber\\
&& \qquad -2 \pi G^2 m_1 m_2 \frac{\gamma}{b_0^2}\frac{(\gamma^2-2)(2\gamma^2-1)}{\gamma^2-1} \nonumber\\
&& \int_{-\infty}^\infty \Gamma({\sf h}_1^{\delta r_2})^1{}_{2\alpha} u_1^-{}^\alpha d\tau_1  = \nonumber\\
&&\qquad  G^2 m_1 m_2 \frac{\gamma}{b_0^2} \left[\pi  (2\gamma^2+1)-4\frac{2\gamma^2-1}{\gamma^2-1} \right] \,.
\end{eqnarray}
Then, the last two terms, quadratic-in-${\sf h}_1$,  in Eq. \eqref{Gamma_h1_all}, namely
\beq
\frac12 \left({\sf h}_1^{yy}-{\sf h}_1^{tt}  \right)\partial_x {\sf h}_1^{ty}-\frac14 {\sf h}_1^{ty} \left(\partial_x {\sf h}_1^{tt}-\partial_x {\sf h}_1^{yy} \right)\,,
\eeq
are found to contribute
\beq
\int_{-\infty}^\infty \Gamma({\sf h}^{0,\rm quad}_1){}_{2\alpha} u_1^\alpha d\tau_1  =-6 G^2 m_2^2 \pi \frac{\gamma}{b_0^2}\,.
\eeq

The last $O(G^2)$ correction comes from  the quadratic contribution in the Dyson expansion, as written in Eq. \eqref{dysonquad}.
This term  uses only ${\sf h}^{0,\rm lin}_1$ and  is found to yield the following contribution to $\Lambda_{(1)}^1{}_2$:
\beq
-\frac12 G^2 m_2^2 \pi \frac{\gamma}{b_0^2}\left[4+\frac{1}{\gamma^2-1}  \right]\,.
\eeq
As a check on our results, we have recomputed all the terms deriving from ${\sf h}_1$ in $\Lambda_{(1)}^1{}_2$  in a completely 
different way, namely by using the Fourier space approach outlined  in Ref. \cite{Damour:2016gwp}. We found complete agreement 
with those obtained above. Some results of the  Fourier space approach are summarized in Appendix A.

\subsection{Final result  for the holonomy map}

Summing up all  the contributions above, and reexpressing the result in terms of the
physical impact parameter $b$ instead of $b_0$ (using Eq. \eqref{bvsb0}), we get the final 2PM-accurate result
\beq
\Lambda_{(1)}^1{}_2=\frac{4Gm_2\gamma}{b}+\frac{G^2\pi m_2\gamma (5\gamma^2-3)}{(\gamma^2-1)b^2}\left(\frac34 m_2+m_1  \right)\,.
\eeq
It will be most useful to rewrite this result in terms of the total orbital c.m. angular momentum
\beq
L=b \, p_{\rm cm}=b m_1 \sinh \alpha\,.
\eeq
We introduce the rescaled version $j$ of $L$ defined by
\beq
L=Gm_1m_2 j\,
\eeq
and the relation
\beq
b  =  \frac{G (m_1+m_2) h j}{\sqrt{\gamma^2-1}}\,,
\eeq
where $h$ denotes
\beq
h \equiv \frac{{\mathcal E}_{\rm real}}{M} =\frac{{\mathcal E}_1+{\mathcal E}_2}{m_1+m_2}= \sqrt{1+ 2 \nu (\gamma-1)}\,.
\eeq
This yields (modulo an $O(1/j^3)$ error term)
\beq
\Lambda_{(1)}^1{}_2 = \frac{4\gamma \sqrt{\gamma^2-1}}{jh}X_2
+\frac{\pi \gamma (5\gamma^2-3)}{j^2h^2}X_2 \left(X_1 +\frac34 X_2 \right)
\eeq

Using then the relation \eqref{lambdavstheta}, i.e.
\beq
\theta_1 = (1-\cosh \alpha)\chi +\Lambda_{(1)}^1{}_2 + O(G^3)\,,
\eeq
(where we have added a label 1 to $\theta$ as a reminder that this is the spin rotation along $\mathcal{L}_1$), with
\beq
\cosh \alpha = \frac{X_1}{h}+\frac{\gamma X_2}{h}\,,
\eeq
as well as the known  2PM expression of the orbital scattering angle $\chi$ \cite{Westpfahl:1985}
\beq
\chi = \frac{2(2\gamma^2-1)}{j\sqrt{\gamma^2-1}}+\frac34 \frac{\pi}{j^2}  \frac{5\gamma^2-1}{h},
\eeq
we finally obtain, at the 2PM level, the angle of rotation $\theta_1$ entering the spin holonomy along $\mathcal{L}_1$:
\begin{eqnarray} \label{thetab}
\theta_1 &=& \frac{2G}{b(\gamma^2-1)}\left[(2\gamma^2-1)(h-1)m_1\right. \nonumber\\
&& \left. +[(2\gamma^2-1)h-\gamma]m_2  \right]\nonumber\\
&&+\frac{G^2\pi }{4 b^2 (\gamma^2-1)} \left[3(5\gamma^2-1)(h-1)m_1^2\right. \nonumber\\
&& +[3(5\gamma^2-1)(2h-1) \nonumber\\
&& +\gamma(5\gamma^2-9)]m_1m_2\nonumber\\
&&\left.  +3[h(5\gamma^2-1)-2\gamma]m_2^2  \right]\,,
\end{eqnarray}
or
\begin{eqnarray} \label{thetaj}
\theta_1 &=& -\frac{2}{hj \sqrt{\gamma^2-1}} \left[ \gamma X_2 +(2\gamma^2-1)(X_1-h) \right]\nonumber\\
&&+\frac{\pi }{4h^2j^2}\left[-3(5\gamma^2-1)(X_1-h)-6\gamma X_2\right.\nonumber\\
&& \left.+\gamma (5\gamma^2-3)X_1X_2  \right]\,.
\end{eqnarray}
Here we have used either the impact parameter $b$ or the dimensionless orbital angular momentum $j$.

\section{EOB computation of the spin-holonomy rotation angle $\theta_1$}

In order to transcribe the  2PM-accurate value \eqref{thetaj} of the spin-holonomy rotation angle $\theta_1$ 
into a corresponding 2PM-accurate value of the two gyrogravitomagnetic factors $g_S$ and $g_{S*}$ parametrizing the
spin-orbit contribution to the effective EOB Hamiltonian, Eq. \eqref{Heff}, we need to compute, within the framework of EOB theory,
the value of $\theta_1$. This was done at the 1PM level in Ref. \cite{Bini:2017xzy}. The basic physical effect 
underlying this computation is conceptually very simple (and was already discussed in Ref. \cite{Bini:2017xzy}). 
Indeed, the EOB-derived Hamiltonian equation of motion for the spin vector 
${\bf S}_1$ of particle 1, as obtained from Eq. \eqref{Heff}, is simply
\beq
\frac{d {\bf S}_1}{dT_{\rm eff}}= {\bf \Omega_1} \times {\bf S}_1
\eeq
where $T_{\rm eff}$ is the effective EOB time (i.e. the evolution parameter with respect to the effective Hamiltonian),
and where
\beq
{\bf \Omega_1}= \frac{\partial H_{\rm eff}}{\partial {\bf S}_1}\,.
\eeq
When considering the linear-in-spin effective EOB Hamiltonian Eq. \eqref{Heff}, the vectorial angular velocity ${\bf \Omega_1}$
is simply given by
\beq
{\bf \Omega_1}= G   \frac{{\mathbf L}}{R^3}\left(g_S +\frac{m_2}{m_1}g_{S*} \right)\,.
\eeq
Moreover, when working linearly in the spins, the orbital angular momentum vector ${\mathbf L}$ can be considered as
being constant. The integrated vectorial angle of rotation of ${\bf S}_1$ is then given by
\begin{eqnarray}
{\boldsymbol \theta}_1^{\rm EOB}&=&G \int {\bf \Omega_1} dT_{\rm eff}\nonumber\\
&=& G \int  \frac{{\mathbf L}}{R^3}\left(g_S +\frac{m_2}{m_1}g_{S*} \right) dT_{\rm eff}.
\end{eqnarray}
To get an explicit integral expression for ${\boldsymbol \theta}_1$ we need to express $dT_{\rm eff}$ in terms of the EOB
radial variable $R$. Such an expression is simply obtained by writing the Hamiltonian evolution equation for $R$,
namely 
\beq \label{eomR}
\frac{d R}{dT_{\rm eff}}= + \frac{\partial H_{\rm eff}}{\partial {P_R}}
\eeq
At linear order in the spins, it is enough to use the orbital part of $H_{\rm eff}$, i.e. (with $L=P_\phi$)
\beq \label{Hefforb}
H_{\rm eff}^{\rm orb}=  \sqrt{A(R)\left( \mu^2  + \frac{P_R^2}{B(R)} + \frac{L^2}{R^2} +Q \right)}\,.
\eeq
In addition, as we work at the 2PM accuracy, and as there is an explicit $G$ factor in the expression of ${\boldsymbol \theta}_1$,
it is enough to use the 1PM-accurate EOB effective Hamiltonian. It was shown in Ref. \cite{Damour:2016gwp} that the
1PM-accurate effective EOB Hamiltonian was simply given by the geodesic dynamics of a particle of mass $\mu$
in a (linearized) Schwarzschild metric of mass $M$, i.e. by an Hamiltonian of the form  \eqref{Hefforb} with
\begin{eqnarray}
A&=&1-2 \frac{GM}{R}+O(G^2) \,, \nonumber\\ 
B&=& 1+2\frac{GM}{R}+O(G^2) \,,\qquad\quad  Q=0 \,.
\end{eqnarray}
Using the equation of motion of $R$, Eq. \eqref{eomR}, to express $dT_{\rm eff}$ in terms of $dR$ then leads to
the following expression for the (vectorial) rotation angle ${\boldsymbol \theta}_1$
\beq
\label{theta_eq}
{\boldsymbol \theta}_1^{\rm EOB}=G \int  \frac{{\mathbf L}}{R^3}\left(g_S +\frac{m_2}{m_1}g_{S*} \right)\frac{B}{A}E_{\rm eff}\frac{dR}{P_R}\,,
\eeq
in which $P_R$ should be expressed as a function of $R$ by using the energy conservation (for zero spins), namely
\beq \label{Eeff}
E_{\rm eff}^2=A \left( \mu^2 +\frac{P_R^2}{B}+\frac{L^2}{R^2}\right)\,,
\eeq
so that
\beq
P_R^2=\frac{B}{A}\left(E_{\rm eff}^2-A\left(\mu^2+\frac{L^2}{R^2}  \right) \right) \,.
\eeq
In the following, we shall consider the (algebraic) scalar magnitude, $\theta_1^{\rm EOB}$, of the spin rotation, such that
${\boldsymbol \theta}_1^{\rm EOB}/{\mathbf L} = \theta_1^{\rm EOB}/L$.

The gyrogravitomagnetic factors $g_S$ and $g_{S*}$ entering Eq. \eqref{theta_eq} are not constant factors but are
functions on  phase-space. In order to be able to explicitly perform the integral over $R$ entering the expression 
\eqref{theta_eq}, it is convenient to choose a spin-gauge which simplifies the phase-space dependence of $g_S$ and $g_{S*}$.
 We recall that Ref. \cite{Damour:2008qf} emphasized the existence
of a rather large gauge freedom in the phase-space dependence of $g_S$ and $g_{S*}$, linked to the freedom in local rotations
of the frame used to measure the local components of the spin vectors ${\bf S}_1$ and ${\bf S}_2$. In our
previous 1PM-level work Ref. \cite{Bini:2017xzy}, we had conveniently used this freedom to express the $G \to 0$ limit
of $g_S$ and $g_{S*}$ as functions of only ${\bf P}^2$, which was conserved at this order. At our present
2PM order it will be convenient to use a spin-gauge such that
\begin{eqnarray} \label{gSSsexp2}
g_{S}&=& g_{S}^{\rm 1PM}(H_{\rm eff}^{\rm orb})+g_{S}^{\rm 2PM}(H_{\rm eff}^{\rm orb}) u  +O(u^2)\,,\nonumber\\
g_{S*}&=& g_{S*}^{\rm 1PM}(H_{\rm eff}^{\rm orb})+g_{S*}^{\rm 2PM}(H_{\rm eff}^{\rm orb}) u +O(u^2)\,.
\end{eqnarray}
As we work linearly in the spins, the use of  $H_{\rm eff}^{\rm orb}$ as argument is equivalent to
using the full $H_{\rm eff}$ as was indicated in Eq. \eqref{gSSsexp1} of the Introduction. The crucial point is that the
value of this argument can be treated as a constant (equal to $E_{\rm eff}$, see Eq. \eqref{Eeff}) during the
integration involved in Eq. \eqref{theta_eq}. In the following, we shall refer to this gauge as the ``energy-gauge".

The integral \eqref{theta_eq} can then be explicitly computed to the required 2PM accuracy by expanding its
integrand in powers of $G$, which means expanding it in powers of
\beq
u=\frac{GM}{R}\,.
\eeq
Let us use the shorthand notation
\begin{eqnarray}
g_{SS*}&\equiv& g_S +\frac{m_2}{m_1}g_{S*}\nonumber\\
&=& g_{SS*}^{\rm 1PM}+g_{SS*}^{\rm 2PM} u + g_{SS*}^{\rm 3PM} u^2 +O(G^3)\,.
\end{eqnarray}
The various $u$- or $G$-expansions we need are
\beq
A=1-2u+O(G^2)\,;\qquad B=1+2u+O(G^2)\,, 
\eeq
so that
\beq
\frac{B}{A}=1+4u+O(G^2)\,,
\eeq
and
\beq
P_R=\pm (P_R^{(0)}+ G P_R^{(1)})\,,
\eeq
with
\begin{eqnarray}
P_R^{(0)}&=&\sqrt{E_{\rm eff}^2-\mu^2 -\frac{L^2}{R^2}}\,,\nonumber\\
G P_R^{(1)}&=& 2u  P_R^{(0)} +u \frac{\mu^2+\frac{L^2}{R^2}}{P_R^{(0)}}\,.
\end{eqnarray}
Substituting in Eq. \eqref{theta_eq} we then formally find 
\begin{eqnarray}
\theta_1^{\rm EOB}&=&  \nonumber\\
&=& 2 G E_{\rm eff} L \int_{R_{\rm min}}^\infty \frac{g_{SS*} }{R^3} \left(1+4 u\right)\times \nonumber\\
&& \times \left(1-G \frac{P_R^{(1)}}{P_R^{(0)}} \right)\frac{dR}{P_R^{(0)}}\nonumber\\
&=& 2 G E_{\rm eff}  \int_{R_{\rm min}}^\infty \frac{g_{SS*} }{R^3}\left(1+4 u-G \frac{P_R^{(1)}}{P_R^{(0)}} \right)\frac{ L dR}{P_R^{(0)}}\,,\nonumber\\
\end{eqnarray}
where the lower limit of the radial integral is
\beq
R_{\rm min}=\frac{L}{\sqrt{E_{\rm eff}^2-\mu^2}}\,.
\eeq
Note that 
\beq
P_R^{(0)}=\sqrt{E_{\rm eff}^2-\mu^2}\sqrt{1-\frac{R_{\rm min}^2}{R^2}} \,,
\eeq
so that $P_R^{(0)}$ vanishes at the lower limit (which, indeed, corresponds to the turning point
of the  $O(G^0)$ radial motion).

A more explicit form of the above integral is then
\begin{widetext}
\beq
\label{theta1_eq_EOB}
\theta_1^{\rm EOB} =2G E_{\rm eff}R_{\rm min}\int_{R_{\rm min}}^\infty  \frac{dR}{R^3\sqrt{1-\frac{R_{\rm min}^2}{R^2}}}\left[
g_{SS*}^{\rm 1PM}+u g_{SS*}^{\rm 2PM}+2u g_{SS*}^{\rm 1PM}-ug_{SS*}^{\rm 1PM} \frac{R_{\rm min}^2}{L^2}\frac{\mu^2+\frac{L^2}{R^2}}{1-\frac{R_{\rm min}^2}{R^2}}
\right]\,.
\eeq
\end{widetext}
This integral is formally divergent at the lower limit, where the last term in the integrand has
a power-law singularity $\propto (1-\frac{R_{\rm min}^2}{R^2})^{-3/2}$. This divergence was absent in
the original, unexpanded form of the integral, and was generated by our formal expansion in powers of $G$.
However, some years ago, Damour and Sch\"afer \cite{Damour:1988mr} have proven a general result about
such singular integrals obtained by formally expanding (in a perturbation parameter) radial integrals involving turning points that are 
perturbed during the formal expansion. The correct result is simply obtained by taking the Hadamard partie finie (PF)
of the formally expanded (singular) radial integrals.

It is technically convenient to use  the new variable 
\beq
x=\frac{R_{\rm min}^2}{R^2}\,,\qquad -\frac12 \frac{dx}{R_{\rm min}^2}=\frac{dR}{R^3}\,,
\eeq
in the integral \eqref{theta_eq}. We then get
\begin{eqnarray}
\theta_1^{\rm EOB} &=& \frac{G E_{\rm eff}}{R_{\rm min}}g_{SS*}^{\rm 1PM}\,\, {\rm PF}\int_0^1 \frac{dx}{\sqrt{1-x}}\nonumber\\
&+ & \frac{G^2 ME_{\rm eff}}{R_{\rm min}^2}\left[\left(g_{SS*}^{\rm 2PM}+2g_{SS*}^{\rm 1PM}-g_{SS*}^{\rm 1PM}\frac{\mu^2}{E_{\rm eff}^2-\mu^2}  \right) \times\right. \nonumber\\
&&\left. \,\, {\rm PF}\int_0^1 \frac{\sqrt{x}dx}{\sqrt{1-x}} \right]\nonumber\\
&& -g_{SS*}^{\rm 1PM} \frac{E_{\rm eff}^2}{E_{\rm eff}^2-\mu^2}\,\, {\rm PF}\int_0^1 \frac{x^{3/2}dx}{(1-x)^{3/2}}\,.
\end{eqnarray}
This result is actually valid in any spin-gauge, but it is particularly useful in what we have called above the energy-gauge, as defined
by Eq. \eqref{gSSsexp2}. Indeed, in this spin-gauge all the coefficients involving 
 $g_{SS*}^{\rm 1PM}$ and $g_{SS*}^{\rm 2PM}$ are constant.
 
The integral can then be easily  obtained, besides using the trivial 1PM-level integral,
\beq
\int_0^1 \frac{dx}{\sqrt{1-x}}=2\,,
\eeq
from the general result:
\beq
{\rm PF}\int_0^1 \frac{x^{(2n+1)/2}dx}{(1-x)^{(2n+1)/2}} =(-1)^n \frac{(2n+1)\pi}{2} \,,\, n=0,1,2,\ldots
\eeq
leading, in particular, to
\begin{eqnarray}
{\rm PF}\int_0^1 \frac{\sqrt{x}dx}{\sqrt{1-x}} &=&\frac{\pi}{2}\,,\nonumber\\
{\rm PF}\int_0^1 \frac{x^{3/2}dx}{(1-x)^{3/2}} &=& -\frac{3\pi}{2}\,.
\end{eqnarray}
The final result of the integration is:
\begin{eqnarray}  \label{thetaEOBfinal}
\theta_1^{\rm EOB} &=&\frac{2\nu}{j}\gamma \sqrt{\gamma^2-1}  g_{SS*}^{\rm 1PM}\\
&+&  \frac{\pi}{2}\frac{\nu}{ j^2}\, \gamma [(5\gamma^2-3)g_{SS*}^{\rm 1PM}+(\gamma^2-1)g_{SS*}^{\rm 2PM}]\,.\nonumber
\end{eqnarray}
Here, we used, as above, the dimensionless orbital angular momentum $j=\frac{L}{GM\mu}$, and we have also 
(to prepare the EOB transcription of the 2PM rotation angle) replaced $E_{\rm eff}$
by
\beq
\gamma =\frac{ E_{\rm eff}}{\mu} \equiv \hat E_{\rm eff}\,.
\eeq
It is indeed a noticeable result of EOB theory that the map between the (squared) real c.m. energy, $s={\mathcal E}_{\rm real}^2$,
Eq. \eqref{s}, and the effective EOB energy $E_{\rm eff}$ is given by the formula (proven in Ref. \cite{Damour:2016gwp} to be 
valid to all orders in $v/c$)
\beq \label{emap1}
{\mathcal E}_{\rm real}=M \sqrt{1+2\nu \left(\frac{E_{\rm eff}}{\mu }-1 \right)}\,,
\eeq
or, equivalently,
\beq \label{emap2}
\frac{ {E}_{\rm eff}}{\mu} =  \frac{({\mathcal E}_{\rm real})^2 - m_1^2  -m_2^2 }{2 m_1 m_2} \,.
\eeq
Inserting $s={\mathcal E}_{\rm real}^2  =  - (p_1^- + p_2^-)^2$ in the latter expression, then yields
\beq \label{emap3}
\frac{ {E}_{\rm eff}}{\mu} =  - \frac{p_1^-  \cdot p_2^- }{ m_1 m_2} = \gamma\,.
\eeq

\section{EOB transcription of the spin-holonomy rotation angle $\theta_1$}

As discussed in    Ref. \cite{Bini:2017xzy}, the basic condition allowing one to transcribe our 2PM-accurate computation
in Sec. \ref{sec5} of the real, two-body spin-rotation angle $\theta_1=\theta_1^{\rm real}$ into information about the EOB spin-orbit couplings
is simply the identification
\beq \label{identif}
\theta_1^{\rm real}(\gamma, L^{\rm real}, m_1, m_2) = \theta_1^{\rm EOB}( {E}_{\rm eff}, L^{\rm EOB},m_1,m_2)\,.
\eeq
Crucial to this identification is the knowledge of the connection between the real dynamical variables (notably $\gamma$, 
and $L^{\rm real}$) appearing as arguments on the left-hand side, and the effective dynamical variables 
(${E}_{\rm eff}, L^{\rm EOB}$) appearing on the right-hand side. On the one hand, one of the basic principles of
EOB theory is that $L^{\rm real}= L^{\rm EOB}$, or $j^{\rm real}= j^{\rm EOB}$. On the other hand, we have already explained above the link predicted
by EOB theory between the real (incoming) Lorentz factor $\gamma$ and the EOB effective energy ${E}_{\rm eff}$.
See, Eq. \eqref{emap3}, which was already used in re-expressing our final result for $\theta_1^{\rm EOB}$.

We note also that, in our energy-spin-gauge, $g_{S}^{\rm 1PM}$,  $g_{S*}^{\rm 1PM}$,  $g_{S}^{\rm 2PM}$,  
and $g_{S*}^{\rm 2PM}$ must all be functions of only $\gamma= \hat E_{\rm eff}$ and of the {\it symmetric} mass ratio $\nu$.

As both sides of the basic identification \eqref{identif} are truncated power series in $1/j$, we get a separate condition at
each PM order. This condition was already solved  at the 1PM order in Ref. \cite{Bini:2017xzy}, with the result (recalling
the notation $X_1\equiv m_1/M$,  $X_2\equiv m_2/M = 1-X_1$)
\begin{eqnarray}
g_{SS*}^{\rm 1PM}(\gamma,X_1)= g_{S}^{\rm 1PM}(\gamma, \nu)+\frac{X_2}{X_1} g_{S* }^{\rm 1PM}(\gamma, \nu)\,,
\end{eqnarray}
with
\begin{eqnarray} \label{gSS1PM}
g_{S}^{\rm 1PM}(\gamma, \nu)&=& \frac{ (2\gamma+1) (2\gamma+h)-1 }{h (h+1) \gamma (\gamma +1) } \nonumber\\
&=& \frac{1}{h(h+1)}\left[4+\frac{h-1}{\gamma+1}+\frac{h-1}{\gamma}  \right]\nonumber\\
g_{S* }^{\rm 1PM}(\gamma, \nu)&=& \frac{ 2\gamma+1 }{ h \gamma(\gamma +1)} \nonumber\\
&=&  \frac{1}{h}\left[ \frac{1}{\gamma+1}+\frac{1}{\gamma} \right]\,.
\end{eqnarray}
The $\nu$-dependence of $g_{S}^{\rm 1PM}$ and  $g_{S*}^{\rm 1PM}$ 
is entirely contained in the quantity $h$. We recall that $h \equiv{\mathcal E}_{\rm real}/M$ 
is the following function of $\gamma = {E}_{\rm eff}/\mu$ and $\nu$:
\beq
h= \sqrt{1+2\nu (\gamma-1)}\,.
\eeq

At the 2PM order, the condition \eqref{identif} gives one equation relating the combination
\beq
g_{SS*}^{\rm 2PM}(\gamma, X_1)=g_{S}^{\rm 2PM}(\gamma, \nu)+\frac{X_2}{X_1}g_{S*}^{\rm 2PM}(\gamma, \nu)\,,
\eeq
to some given function of $\gamma$, and $X_1$, say $f(\gamma,X_1)$. 
One can extract the two independent (but symmetric under the $1 \leftrightarrow 2$ exchange) functions $g_{S}^{\rm 2PM}(\gamma, \nu)$ and $g_{S*}^{\rm 2PM}(\gamma, \nu)$
from the single condition
\beq
g_{S}^{\rm 2PM}(\gamma, \nu)+\frac{X_2}{X_1}g_{S*}^{\rm 2PM}(\gamma, \nu) = f(\gamma,X_1).
\eeq
Indeed, the right-hand side of this equation is a dissymetric function of the two masses.
By considering suitable symmetric combinations (such as $ f(\gamma,X_1)+  f(\gamma,X_2)$
and $  [f(\gamma,X_1)-  f(\gamma,X_2)]/(X_1-X_2)$) one gets two independent equations allowing
one to determine the two symmetric functions $g_{S}^{\rm 2PM}(\gamma, \nu)$ and $g_{S*}^{\rm 2PM}(\gamma, \nu)$.
[We recall that $\nu \equiv X_1 X_2$.]

By equating our two 2PM-accurate results \eqref{thetaj} and \eqref{thetaEOBfinal} we finally get the 2PM-accurate values
of the two gyrogravitomagnetic ratios
\begin{widetext}
\begin{eqnarray} \label{gSS2PM}
g_{S}^{\rm 2PM}(\gamma, \nu)&=& -\frac{\nu}{\gamma (\gamma+1)^2 h^2 (h+1)^2}[2(2\gamma+1)(5\gamma^2-3)h+ (\gamma+1)(35\gamma^3-15 \gamma^2-15\gamma +3)] \nonumber\\
&=& \frac{\nu}{h^2(h+1)^2}\left[ -5 (7\gamma +4h-10)+\frac{8 (3h-4)}{\gamma+1}-\frac{4  h}{(\gamma+1)^2}+\frac{3  (2h-1)}{\gamma} \right]\nonumber\\
g_{S* }^{\rm 2PM}(\gamma, \nu)&=& -\frac{1}{2\gamma (\gamma+1)^2 h^2 (h+1)}\left[ (5\gamma^2+6\gamma+3)(h+1)+4\nu(1+2\gamma)(5\gamma^2-3)  \right]\nonumber\\
&=& \frac{1}{h^2(h+1)}\left[ -20\nu +\frac{24\nu -h-1}{\gamma +1} +\frac{h+1-4\nu}{(\gamma +1)^2}-\frac{3}{2}\frac{h+1-4\nu}{\gamma}\right]\nonumber\\
&=& \frac{1}{h^2(h+1)}\left[-\frac{20\gamma \nu}{\gamma+1}+(h+1-4\nu)\left( \frac{1}{(\gamma+1)^2}-\frac{1}{\gamma+1}-\frac{3}{2}\frac{1}{\gamma} \right) \right] \,.
\end{eqnarray}
\end{widetext}
These are the two main results of the present work.

\section{Comparison with PN-expanded results}

One of the crucial checks on our 2PM-accurate computation of the gyrogravitomagnetic ratios is their
comparison with the known sub-sub-leading order PN-expanded values of these ratios. [This check has helped us
to better locate  the subtle aspects of the 2PM calculations explained in Sec. \ref{sec5} above.]
In the present section we restore the presence of the velocity of light $c$ in formulas.

We recall that the PN-expansions of $g_S$ and $g_{S*}$ were derived at the next-to-leading order (or sub-leading order) in
Ref. \cite{Damour:2008qf}, and at the sub-sub-leading order in Refs. \cite{Nagar:2011fx} and \cite{Barausse:2011ys}.
These determinations were done in a general spin gauge, i.e. allowing for a large arbitrariness in the rotational state
of the local spin frame. 

We have determined the values of the spin-gauge parameters corresponding to the energy-spin-gauge
we found most useful in our 2PM calculation. To do so, it is useful to replace the dimensionless effective 
energy $\hat E_{\rm eff} = E_{\rm eff}/ (\mu c^2)$, or rather the dimensionless effective Hamiltonian
$\hat H_{\rm eff} = H_{\rm eff}/ (\mu c^2)$, by the phase-space variable $W({\bf Q}, {\bf P})$, such that
\beq
 \hat H_{\rm eff}^2({\bf Q}, {\bf P}) \equiv 1+ \frac{W({\bf Q}, {\bf P})}{c^2}\,.
\eeq
In the PN expansion, one considers that $1/c\to 0$, keeping fixed $W$ as well as the rescaled momenta ${\bf p} \equiv {\bf P}/\mu$
and ${ p}_r \equiv { P}_R/\mu$. 
To sufficient accuracy, we can use the effective Hamiltonian corresponding to an effective Schwarzschild metric (with $u=GM/R$)
\beq
 \hat H_{\rm eff} = \sqrt{\left(1-2 \frac{u}{c^2}\right) \left(1+\left(1-2 \frac{u}{c^2}\right) \frac{ p_r^2}{c^2}+\frac{p^2}{c^2}- \frac{p_r^2}{c^2}\right)}\,. 
\eeq
[At the present, sub-sub-leading order, we do not need 2PN-level corrections to the effective metric.]

We can then use the previous relations to express $p$ in terms of $W$ and $p_r$:
\begin{eqnarray}
p &=& \sqrt{W+2 u}+\frac{1}{c^2}\frac{ (2 u^2+ p_r^2 u+ W u)}{ (W+2 u)^{1/2} }\\
&+& \frac{1}{c^4} \frac{ u^2 [-p_r^4-2(  W+2 u) p_r^2+3 W^2+12 W u+12 u^2]}{2(W+2 u)^{3/2}}  \,.\nonumber
\end{eqnarray}
Substituting $p^2$ into the general gauge-dependent expressions of $g_S(p^2,p_r^2,u)$ and $g_{S*}(p^2,p_r^2,u)$ 
derived in \cite{Nagar:2011fx} allows us to express $g_S$ and $g_{S*}$ as functions of the three variables
$W$, $p_r^2$ and $u$. We can then look for values of the spin-gauge parameters that completely remove  all $p_r^2$  terms.
We found a unique set of such gauge-parameters, namely (in the notation of \cite{Nagar:2011fx})
\begin{eqnarray}
&& a = -\frac{3}{2}\nu,\quad b = -\frac{5}{4}\nu,\quad \eta = \frac{9}{16}\nu+\frac{5}{8}\nu^2,\nonumber\\
&& \alpha = -\frac{5}{8}\nu-\frac{7}{4}\nu^2,\quad  \beta  = \frac{27}{16}\nu^2,\nonumber\\
&& \delta = \frac{7}{8}\nu-\frac{5}{4}\nu^2,\quad \gamma  = \frac{7}{8}\nu^2,\quad \zeta = \frac{3}{2}\nu^2\,.
\end{eqnarray}

This allows us to compute the sub-sub-leading PN-expanded values of $g_S$ and $g_{S*}$ in the energy-gauge, i.e. as
functions of $W$ and $u$
\begin{eqnarray}
g_S(W,u) &=& 2+\frac1{c^2}\left(-\frac{9}{8}\nu W-\frac{7}{4} u\nu\right) \nonumber\\
&& +\frac{1}{c^4}\left[\left(-\frac{129}{8}\nu+\frac{5}{4}\nu^2\right) u^2\right.\nonumber\\
&&
+\left(\frac{9}{4}\nu^2-\frac{35}{8}\nu\right) W u \nonumber\\
&& \left. +\left(\frac{1}{8}\nu+\frac{7}{8}\nu^2\right) W^2  \right]\nonumber\\
g_{S*}(W,u) &=& \frac32 +\frac1{c^2}\left[-\left(\frac{5}{8}+\frac{3}{4}\nu\right) W-\left(\frac{7}{4}+\frac{3}{2}\nu\right) u\right]\nonumber\\
&&+\frac{1}{c^4}\left[\left(\frac{7}{16}+\frac12 \nu+\frac{9}{16}\nu^2\right) W^2\right. \nonumber\\
&& +\left(\frac{3}{4}-\nu+\frac{15}{8}\nu^2\right) u W\nonumber\\
&& \left. +\left(-\frac{3}{4}-\frac{21}{2}\nu+\frac{9}{8}\nu^2\right) u^2\right]\,.
\end{eqnarray}

We can then re-order these PN expansions according to powers of $u=GM/R$, thereby determining
the PN-expanded versions of the coefficients of the PM expansion of $g_S$ and $g_{S*}$, namely
\begin{eqnarray}
g_S &=& g_S^{\rm 1PM}(\hat W) +g_S^{\rm 2PM}(\hat  W) \frac{u}{c^2} +g_S^{\rm 3PM}(\hat  W) \frac{u^2}{c^4}\,,\nonumber\\
g_{S*} &=&  g_{S*}^{\rm 1PM}(\hat  W) +g_{S*}^{\rm 2PM}(\hat  W) \frac{u}{c^2} + g_{S*}^{\rm 3PM}(\hat  W) \frac{u^2}{c^4}\,,
\end{eqnarray}
where $\hat W=W/c^2$ and
\begin{eqnarray}
g_S^{\rm 1PM}(\hat W)&=& 2-\frac{9}{8}\nu \hat W   +\left(\frac{1}{8}\nu+\frac{7}{8}\nu^2\right) \hat W^2 +O(\hat W^3)\nonumber\\
g_S^{\rm 2PM}(\hat W)&=&  -\frac{7}{4}  \nu   +\left(\frac{9}{4}\nu^2-\frac{35}{8}\nu\right)\hat W  +O(\hat W^2) \nonumber\\ 
g_S^{\rm 3PM}(\hat W) &=&   -\frac{129}{8}\nu+\frac{5}{4}\nu^2  +O(\hat W)  \nonumber\\
g_{S*}^{\rm 1PM}(\hat W)&=& \frac32 -\left(\frac{5}{8}+\frac{3}{4}\nu\right) \hat W  \nonumber\\
&& + \left(\frac{7}{16}+\frac12 \nu+\frac{9}{16}\nu^2\right) \hat W^2 +O(\hat W^3)\nonumber\\
g_{S*}^{\rm 2PM}(\hat W)&=& -\left(\frac{7}{4}+\frac{3}{2}\nu\right)  + \left(\frac{3}{4}-\nu+\frac{15}{8}\nu^2\right) \hat W\nonumber\\
&& +O(\hat W^2)\nonumber\\
g_{S*}^{\rm 3PM}(\hat W)&=&  -\frac{3}{4}-\frac{21}{2}\nu+\frac{9}{8}\nu^2 +O(\hat W) \,.
\end{eqnarray}

We have checked that the PN expansion (i.e. the expansion in powers of $\hat W$) of our  1PM-level, and 2PM-level,
results, Eqs. \eqref{gSS1PM}, \eqref{gSS2PM}, fully agree with all the corresponding terms in the PN expansions above.

We note that our new results allow one to replace the current, limited-accuracy PN-expanded versions of the four quantitities
$g_{S}^{\rm 1PM}, g_{S*}^{\rm 1PM}, g_{S}^{\rm 2PM}, g_{S*}^{\rm 2PM}$ by four, exactly known functions of $\hat W$.
On the other hand, the only knowledge we currently have concerning the 3PM level is embodied in the two numbers
\beq \label{gS3pm}
\lim_{W \to 0}g_S^{\rm 3PM}( W) =   -\frac{129}{8}\nu+\frac{5}{4}\nu^2\,,
\eeq
and
\beq \label{gSs3pm}
\lim_{W \to 0}g_{S*}^{\rm 3PM}( W)=  -\frac{3}{4}-\frac{21}{2}\nu+\frac{9}{8}\nu^2 \,.
\eeq

\section{High-energy (HE) behavior, strong-field behavior and resummation}

\subsection{High-energy (HE) behavior}

One of the great interests in replacing PN-expanded results by PM-extended ones is that it allows one to
discuss  regimes of the gravitational interaction involving high kinetic energies. Examples of the new insights
obtained this way have been discussed in Refs. \cite{Damour:2016gwp}, \cite{Bini:2017xzy}
and \cite{Damour:2017zjx}. In particular, an interesting property of the high-energy behavior
of the spin-orbit sector has been pointed out in \cite{Bini:2017xzy}. Namely, the fact that, at the 1PM level,
$g_{S}^{\rm 1PM}(\gamma, \nu)$, and $g_{S*}^{\rm 1PM}(\gamma, \nu)$ tend to zero at high-energy (HE),
in specific ways, that differ for $g_{S}$ and  $g_{S*}$:
\beq
g_{S}^{\rm 1PM}(\gamma,\nu) \sim \frac{1}{\nu \gamma }  \; {\rm as} \; \gamma \to \infty\,,
\eeq
and
\beq
g_{S*}^{\rm 1PM}(\gamma,\nu) \sim \frac{1}{ \gamma \sqrt{2 \nu \gamma}}  \; {\rm as} \; \gamma \to \infty\,.
\eeq
It is noticeable that these HE limiting behaviors still hold at the 2PM level, so that it is tempting to
conjecture that they hold at all PM levels. More precisely, we find that, in the HE limit $\gamma \to \infty$, we have
\begin{enumerate}
  \item 
  \begin{eqnarray*}
   g_S(\gamma,u,\nu)&=&g_{S}^{\rm 1PM}(\gamma,\nu)+u\, g_{S}^{\rm 2PM}(\gamma,\nu)\nonumber\\
& \to& 2 \left(1-\frac{35}{8}u\right)\frac{1}{\nu \gamma } +O\left(\frac{1}{\gamma^{3/2}} \right) ;
  \end{eqnarray*}
  \item 
  \begin{eqnarray*}
g_{S*}(\gamma,u,\nu)&=&g_{S*}^{\rm 1PM}(\gamma,\nu)+u \, g_{S*}^{\rm 2PM}(\gamma,\nu)\nonumber\\ 
&\to& 2(1-5u) \frac{1}{\gamma \sqrt{2 \nu \gamma}}  +O\left(\frac{1}{\gamma^{2}} \right)\,.
\end{eqnarray*}
\end{enumerate}
We conjecture that we would have more generally, as $\gamma \to \infty$,
 \begin{eqnarray} 
\label{HEgS}
  g_S(\gamma,u,\nu) &\to & f_S(u) \frac{1}{\nu \gamma } +O\left(\frac{1}{\gamma^{3/2}} \right);\\
\label{HEgSs}
  g_{S*}(\gamma,u,\nu) &\to& f_{S*}(u) \frac{1}{\gamma \sqrt{2 \nu \gamma}}  +O\left(\frac{1}{\gamma^{2}} \right);
\end{eqnarray}
with
\begin{eqnarray} \label{fS}
f_S(u)&=& 2 \left(1-\frac{35}{8}u\right) + O(u^2);\\
f_{S*}(u)&=& 2 \left(1- 5 u\right) + O(u^2)\,.
\end{eqnarray}

There are several interesting consequences of these HE behaviors.
A consequence concerns the HE behavior of the spin-rotation angle $\theta_1$.
Let us first recall that the study of the HE limit of the EOB Hamiltonian \cite{Damour:2017zjx}  has shown 
that it could still be parametrized by an effective metric, with some ($\nu$-independent!) values of the metric functions
$A(R)$ and $B(R)$. [Actually, only the conformal structure of this effective HE metric matters in the HE limit.
It is convenient to fix the conformal freedom by using a Schwarzschild-like (areal radius) gauge, and
we will do so.] 
Using then the EOB-predicted integral expression for $\theta_1$, Eq. \eqref{theta_eq} (with these
effective values of $A(R)$ and $B(R)$), and inserting in this integral the limiting behaviors \eqref{HEgS}, \eqref{HEgSs},
we first notice that the faster decrease of $g_{S*}$ at HE implies the relative disappearance of the
contribution $\propto \frac{m_2}{m_1}g_{S*}$. This immediately shows that the spin rotation angle $\theta_1$
along $\mathcal{L}_1$ will be equal to the spin rotation angle $\theta_2$ along $\mathcal{L}_2$. Taking into account
the various factors of $\nu$ entering into the HE limit of Eq. \eqref{theta_eq}, one also finds
that the common value, say $\theta^{\rm spin}$ of $\theta_1$ and $\theta_2$ 
has a finite value in the HE limit, which is independent of $\nu$. 

Ref. \cite{Damour:2017zjx} found that  the
HE limit of the orbital scattering angle is given by 
\beq \label{chigen2HE}
\frac{\pi}{2}+\frac{\chi}{2} \overset{\rm HE}{=} \int_0^{u_{\rm max}(\bar \alpha)} \!\!du \frac{\sqrt{A(u) B(u)} }{\sqrt{\bar\alpha^2 - u^2 A(u) } }\,,
\eeq
where $u_{\rm max}(\bar \alpha)$ is the root of radicand closest to zero, and where we have set
\beq \label{defalpha}
\bar\alpha \equiv \frac{{\hat E}_{\rm eff}}{j} \equiv \frac{G M E_{\rm eff}}{L}\,.
\eeq
The quantity $\bar\alpha$ is kept fixed in the HE limit where $E_{\rm eff}$ and $j$ both tend to infinity. 

The corresponding HE limit of $\theta^{\rm spin}$ is then found to be given by the integral  %{\bf CHECK}
\beq \label{HEtheta}
 \frac{\theta^{\rm spin}}{2} \overset{\rm HE}{=}\int_0^{u_{\rm max}(\bar \alpha)} \!\! u \, du \, f_S(u)  \sqrt{\frac{B(u)}{A(u)}}\frac{1}{\sqrt{\bar \alpha^2- u^2A(u) }}
 \eeq
The first point we wish to emphasize is that the existence of a finite (mass-independent) HE limit for $\theta^{\rm spin}$ has
directly followed (within the other tenets of EOB theory) from the special ($\nu$-dependent!) HE asymptotic
behavior \eqref{HEgS}. The second point is that the actual value of the limiting spin-rotation angle $\theta^{\rm spin}$
is directly related to the value of the renormalized HE gyrogravitomagnetic ratio $f_S(u)$.
When using the 1PM value of $f_S(u)$, Eq. \eqref{fS}, the computation of the integral \eqref{HEtheta} yields
\beq
\theta^{\rm spin } \overset{\rm HE}{=} 4\bar \alpha  + \frac58\pi \bar \alpha^2 +O(\bar \alpha^3)\,.
\eeq
This value can also, evidently, be directly obtained from taking the corresponding limit in the expression \eqref{thetaj}
The corresponding HE value of the orbital scattering angle is \cite{Damour:2017zjx}
\beq
\chi  \overset{\rm HE}{=} 4\bar \alpha + O(\bar \alpha^3)\,.
\eeq
It is interesting to note that, though these values are independent of the mass ratio, the corresponding HE values of
$\theta^{\rm spin }$ and $\chi$ in the test-particle limit (i.e. when taking $\nu \to 0$ {\it before} taking the HE limit)
are different from the above results, but agree among themselves. 

Let us first recall the value of $g_{S*}$ for a test particle moving in a spherically symmetric metric 
\cite{Barausse:2009aa,Barausse:2009xi,Bini:2015xua}
\beq
g_{S*}^{\rm test}=\frac{r^2\nabla \sqrt{A}}{1+\sqrt{K}}+\frac{r(1-\nabla r)\sqrt{A}}{\sqrt{K}}
\eeq
where $K \equiv {\hat H}_{\rm eff}^2/A$ and $\nabla\equiv B^{-1/2} d/dr$.
For the Schwarzschild values of $A$ and $B$ one finds
\beq
\label{g_Sstar_test}
g_{S*}^{\rm test}=  \frac{\sqrt{1-2u}}{\hat E_{\rm eff}+\sqrt{1-2u}}+\frac{1-2u}{u \hat E_{\rm eff}}\left(1-\sqrt{1-2u} \right).
\eeq
Contrary to the previous case where it was $g_S$ which was dominating in the HE limit, when the test-particle limit is done
before considering the HE limit, it is $g_{S*}$ which dominates. One then finds 
\beq
\frac{\theta^{\rm test}}{2} \overset{\rm HE}{=}   \int_0^{u_{\rm max}(\bar \alpha)}    du \,\, \left(1-\frac{ 1-3u }{\sqrt{1-2u} } \right)\,\, \frac{1}{\sqrt{\bar \alpha^2-u^2(1-2u)}}
\,.
\eeq

A direct computation shows that
\beq
\int_0^{u_{\rm max}(\bar \alpha)}    du \,\,  \frac{ 1-3u }{\sqrt{1-2u} } \,\, \frac{1}{\sqrt{\bar \alpha^2-u^2(1-2u)}}=\frac{\pi}{2}
\eeq
independent of $\bar \alpha$, so that
\beq
\theta^{\rm test}  \overset{\rm HE}{=} \chi^{\rm test}\,,
\eeq
where $\chi^{\rm test}$ is the test-particle limit of the HE orbital scattering angle, given by:
\beq 
\frac{\pi}{2}+\frac{\chi^{\rm test}}{2} \overset{\rm HE}{=} \int_0^{u_{\rm max}(\bar \alpha)} \!\!du \frac{1 }{\sqrt{\bar\alpha^2 - u^2 A(u) } }\,,
\eeq
The beginning of the HE expansions of  $\theta^{\rm spin }$ and $\chi$ read (see \cite{Damour:2017zjx}
for more terms in the expansion)
\beq
\chi^{\rm test} \overset{\rm HE}{=} \theta^{\rm test}_{\rm spin } \overset{\rm HE}{=} 4\bar \alpha+  \frac{15}{4}\pi \bar \alpha^2 +O(\bar \alpha^3)\,.
\eeq

\subsection{Strong-field behavior and resummation}

It was noticed early on \cite{Damour:2008qf} that the PN corrections to the leading PN-order values of $g_S$ and $g_{S*}$,
namely $g^{\rm LO}_S=2$ and $g^{\rm LO}_{S*}= \frac32$ tended (for the comparable-mass case, $4 \nu \sim 1$)
to be all negative, and thereby to diminish the
values of $g_S$ and $g_{S*}$ in the strong-field regime, i.e. when the two bodies get close
to each other so that $u=GM/(c^2R)$ becomes of order unity. In the case of 
$g_{S*}$, its exact test-mass expression \eqref{g_Sstar_test} shows indeed that $g_{S*}$ contains effects that make it
tend to zero both at large energies, and at the horizon $R=2GM/c^2$, i.e. $u=\frac12$. We wish to emphasize here that
the decrease of $g_S$ and $g_{S*}$ as $u$ increases is significantly amplified when considering  large energies.
Indeed, after having factored the overall power-law decreases of $g_S$ and $g_{S*}$ with energy, as in Eqs. \eqref{HEgS}
and \eqref{HEgSs}, we see that the linear slope of fractional decrease of $g_S$ and $g_{S*}$ as $u$ increases becomes
so large (namely $\propto 1-\frac{35}{8}u$ and $\propto 1- 5 u$)  that they would formally predict $g_S$ and $g_{S*}$
to vanish at, respectively, separations $R= \frac{35}{8} GM/c^2= 4.375 GM/c^2$ and $R= 5 GM/c^2$, and then to become
negative. 

To have  $g_S$ and $g_{S*}$ changing sign as the two bodies get close to each other does not {\it a priori} seem
to be physically acceptable. On the other hand, the fact that the HE behavior is factorizable (i.e. seems to be
the same at each PM order) suggests a simple way of resumming the PM expansions of $g_S(\gamma, u)$ and $g_{S*}(\gamma, u)$.
The idea is first to factor out the $\gamma$-dependence that exist at $u=0$, and then to inverse-resum the result
of this factorization. In other words, we suggest to replace the standard PM expansions
\begin{eqnarray}
g_{S,S*}(\gamma, \nu, u)&=& g_{S,S*}^{\rm 1PM}(\gamma, \nu)+u \, g_{S,S*}^{\rm 2PM}(\gamma, \nu)\nonumber\\
&& +u^2 g_{S,S*}^{\rm 3PM}(\gamma, \nu)+O(u^3) 
\end{eqnarray}
by the following resummed expressions
\beq
\label{resum_gsstar}
g_{S,S*}(\gamma, \nu, u)=\frac{g_{S,S*}^{\rm 1PM}(\gamma, \nu)}{1+u \, \tilde  c^1_{S,S*}(\gamma, \nu) +u^2 \, \tilde c^2_{S,S*}(\gamma, \nu)}\,,
\eeq
where
\beq
\tilde  c^1_{S,S*}(\gamma, \nu)= -\frac{g_{S,S*}^{\rm 2PM}(\gamma, \nu)}{g_{S,S*}^{\rm 1PM}(\gamma, \nu)}
\eeq
and  
\beq
\tilde  c^2_{S,S*}(\gamma, \nu)= \left(\frac{g_{S,S*}^{\rm 2PM}(\gamma, \nu)}{g_{S,S*}^{\rm 1PM}(\gamma, \nu)}\right)^2 - \frac{g_{S,S*}^{\rm 3PM}(\gamma, \nu)}{g_{S,S*}^{\rm 1PM}(\gamma, \nu)}\,.
\eeq

\begin{figure*}
\[
\begin{array}{cc}
\includegraphics[scale=0.35]{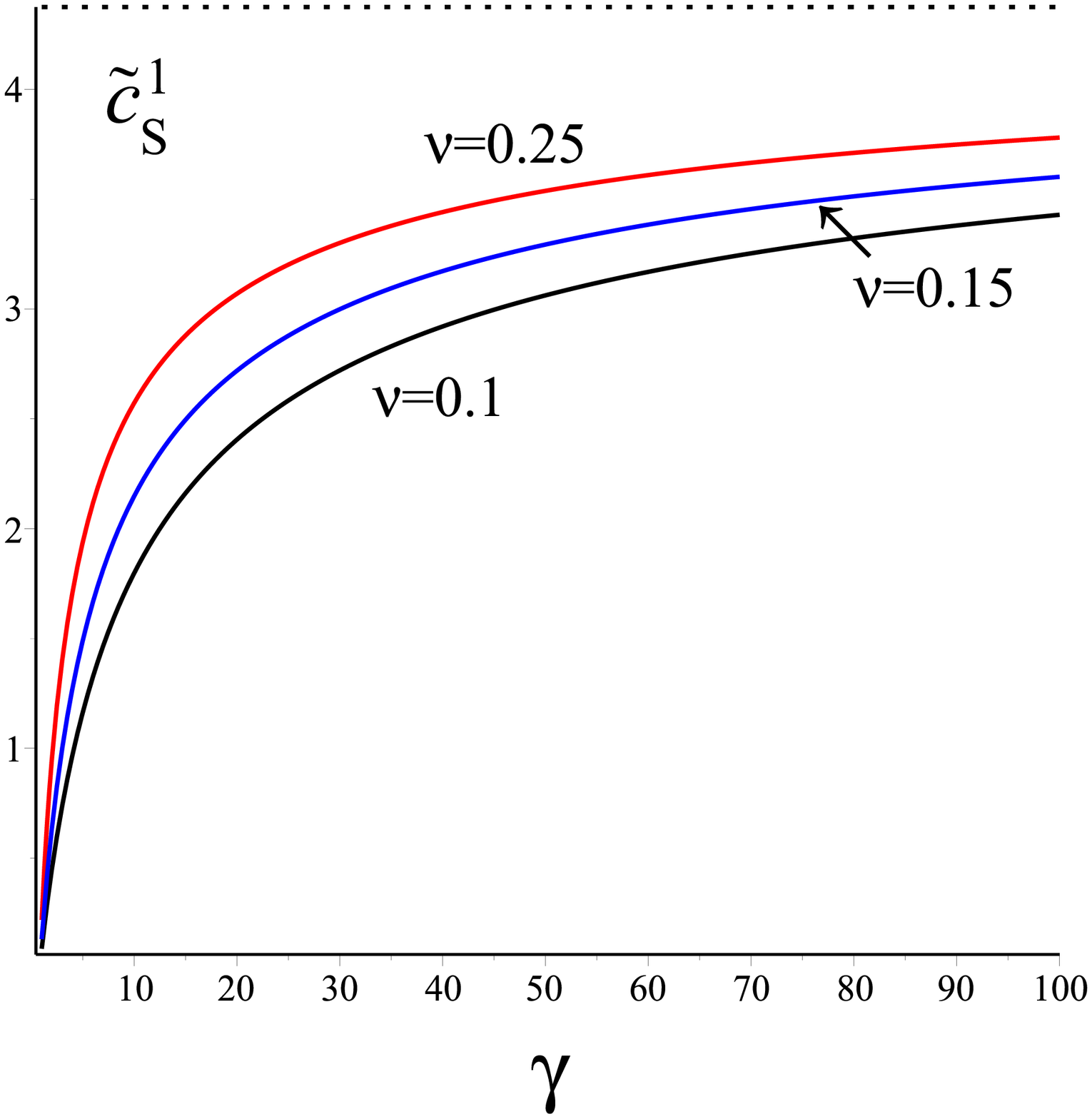} &\qquad
\includegraphics[scale=0.35]{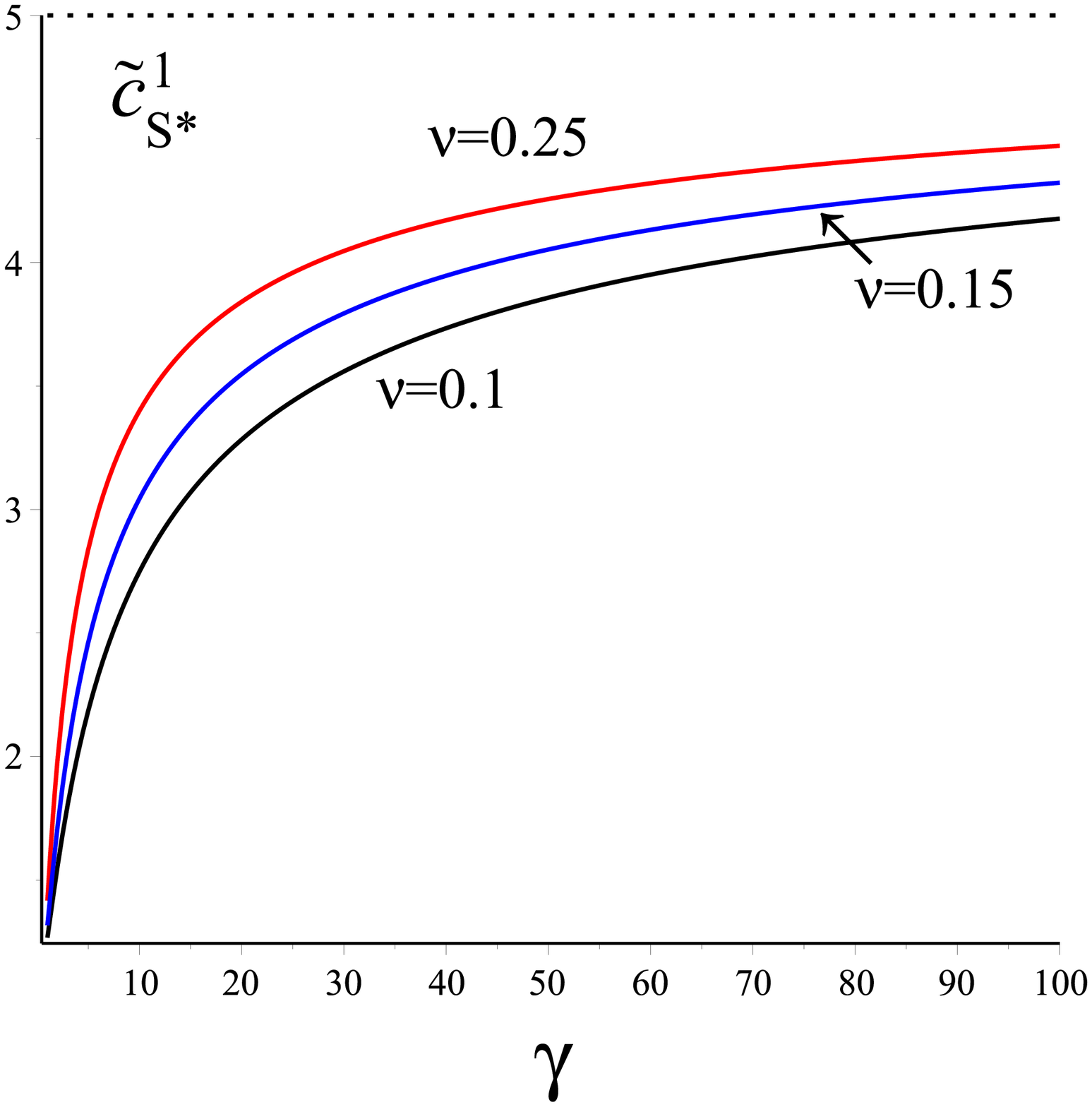}\cr
(a)&
(b)\cr
\end{array}
\]
\caption{\label{fig:tildec}
Panel (a). The behavior of $\tilde c^1_{S}$ is shown as a function of $\gamma$ for selected values of $\nu=[0.1 \,\hbox{(black online)},0.15 \, \hbox{(blue online)},0.25\,  \hbox{(red online)}]$.
The dotted line corresponds to the asymptotic value $\tilde c^1_{S}{}^\infty=35/8$. Panel (b). The behaviour of $\tilde c^1_{S*}$ is shown as for $\tilde c^1_{S}$ with asymptotic value $\tilde c^1_{S*}{}^\infty=5$.}
\end{figure*}

The dependence of the coefficients $\tilde  c^{1}_{S,S*}(\gamma, \nu)$ on $\gamma$ (and the symmetric mass-ratio $\nu$) is shown in Fig. \ref{fig:tildec}.
Both quantities are positive, increasing with $\gamma$ (up to an asymptotic value, $\tilde c^1_{S}{}^\infty=35/8$ and $\tilde c^1_{S*}{}^\infty=5$, respectively) and have a mild dependence on the symmetric mass ratio $\nu$.

Concerning the next coefficients, $\tilde c^2_{S}(\gamma,\nu)$, $\tilde c^2_{S*}(\gamma,\nu)$ we only know their 
low-energy values, i.e. their values for $\gamma=1$. These are easily deduced from the sub-sub-leading results \eqref{gS3pm},
\eqref{gSs3pm} above and read
\begin{eqnarray}
\tilde c^2_{S}(1,\nu) &=& \frac{9}{64}\nu^2+\frac{129}{16}\nu\,,\\
\tilde c^2_{S*}(1,\nu) &=& \frac{7}{16}\nu^2+\frac{91}{12}\nu+\frac{125}{72}\,.  
\end{eqnarray}
In view of the above significant increase of $\tilde c^1_{S}$, $\tilde c^1_{S*}$ as $\gamma$  increases, we might also
expect a significant variation of $\tilde c^2_{S}$, $\tilde c^2_{S*}$ with $\gamma$. We recommend to define
new EOB codes incorporating the
spin-gauge versions of $g_S$ and $g_{S*}$ employed in this paper, together with the inverse-resummed expressions above,
and to compare the predictions of such codes to numerical simulations  of two spinning black holes to try to
determine best-fit values for $\tilde c^2_{S}$, and $\tilde c^2_{S*}$ (and/or for higher-order coefficients).
It would also be interesting to transcribe the high-PN-order results on $g_S$ and $g_{S*}$ obtained (in
a different spin-gauge) from Self-Force theory into the energy-gauge used in this paper.
As, in the context of coalescing binary black holes, the effective energy, $\gamma$, varies numerically little around 1,
it might be sufficient to approximate the rather complicated functions $\tilde c^n_{S}$, $\tilde c^n_{S*}$
by simplified expressions.

\begin{figure*}
\[
\begin{array}{cc}
\includegraphics[scale=0.35]{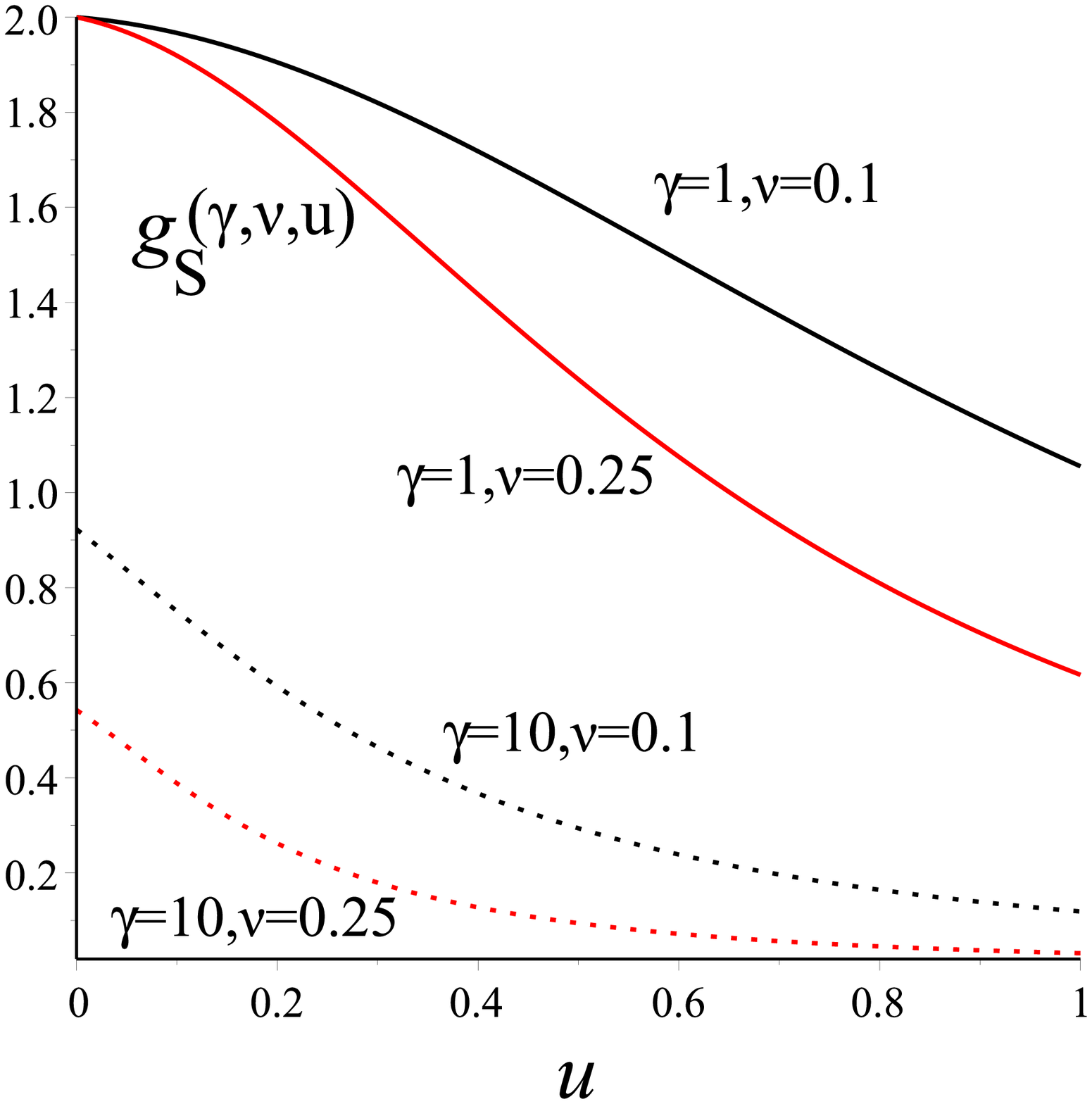} & \qquad
\includegraphics[scale=0.35]{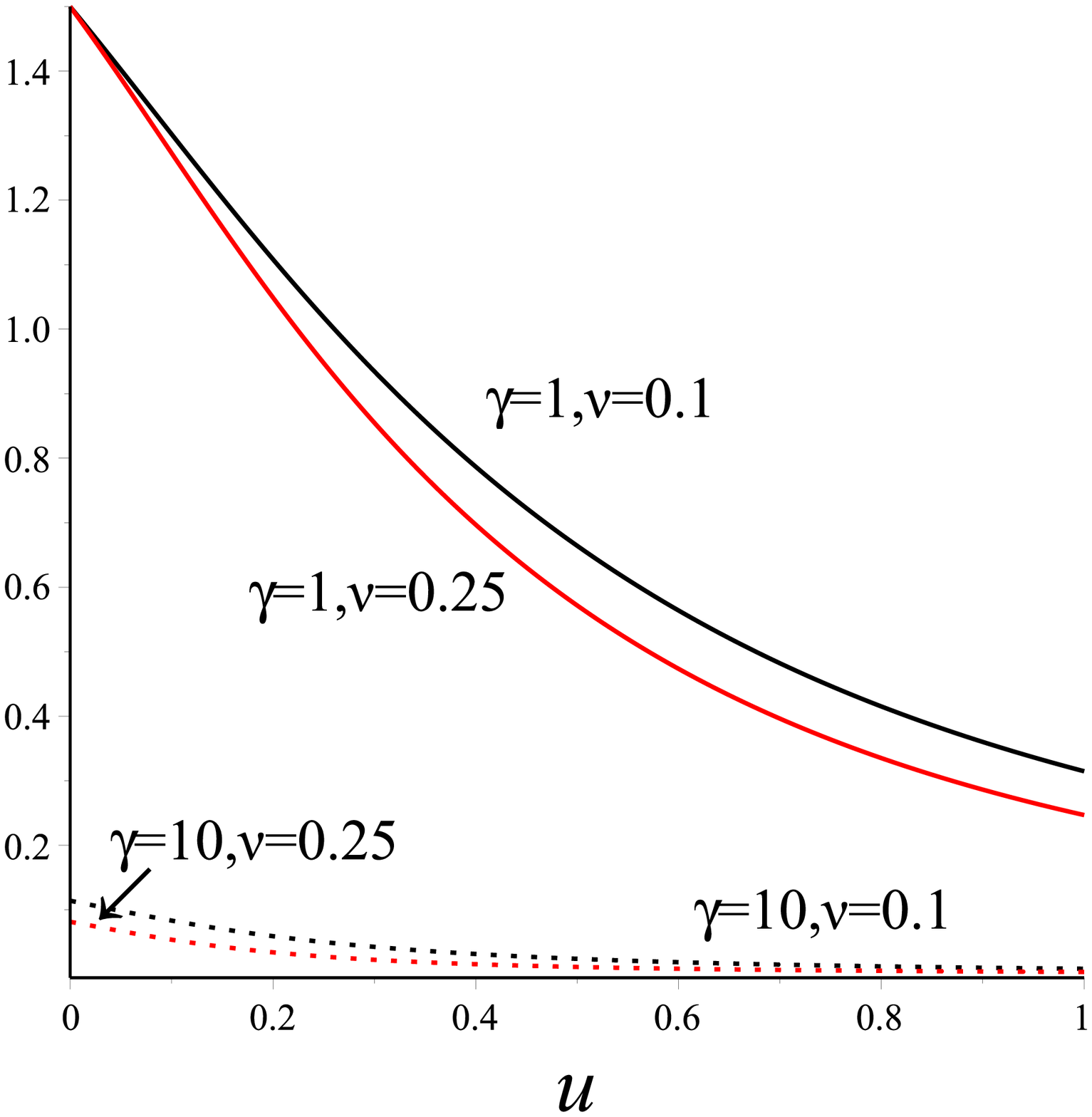}\cr
(a)& (b)\cr
\end{array}
\]
\caption{\label{fig:2} Panel (a). The behavior of $g_S(\gamma,\nu,u)$ as given in Eq. \eqref{resum_gsstar} (with known 3PM terms included) is plotted as a function of $u$ for different values of $\gamma=1,10$ and $\nu=0.1\, \hbox{(black online)},0.25\, \hbox{(red online)}$. Panel (b). $g_{S*}(\gamma,\nu,u)$ is plotted as a function of $u$ for the same parameter choice of  panel (a).}
\end{figure*}

To give an idea of the values of $g_S(\gamma,\nu,u)$ and $g_{S*}(\gamma,\nu,u)$ predicted by our new results, and our new gauge,
we plot in Fig. \ref{fig:2} the $u$-dependences predicted by our formulas (together with the leading-PN order values  of $\tilde c^2_{S}$
and $\tilde c^2_{S*}$) for a sample of values of $\gamma$ and $\nu$. 

We wish also to mention an alternative way of resumming  our 2PM results for the gyrogravitomagnetic ratios $g_{S}$ and $g_{S*}$, 
which is suggested by the (known) structure of the test-particle expression $g_{S*}^{\rm test}$, Eq. \eqref{g_Sstar_test},
for $g_{S*}$ \cite{Barausse:2009aa,Barausse:2009xi,Bini:2015xua}. We see on this expression that the numerators
and denominators of $g_{S*}(\gamma,u)$ contain linear combinations of $\gamma$ with functions of $u$. When expanding
such expressions in powers of $u$ this will generate more complicated $\gamma$-dependent denominators, of the type
seen in our result for $g_{S,S*}^{2 \rm PM}(\gamma)$.  This suggest to resum $g_{S*}$, at the 2PM accuracy, by 
an expression of the following type
\beq \label{gSsresum2}
g_{S*}^{\rm 2PM resum}=\frac{1}{h}\left[ \frac{1+\tilde a_2 u }{\gamma +1+\tilde a_1 u}+\frac{1+\tilde a_3  u}{\gamma}+\tilde a_4 u  \right] \,.
\eeq
Expanding this expression in powers of $u$ and identifying it (at sub-leading order) with our 2PM-accurate result 
determines the various coefficients entering Eq. \eqref{gSsresum2} to be
\beq
\tilde a_i =\frac{a_i}{h(h+1)}\,,
\eeq
with
\begin{eqnarray}
a_1 &=& -h-1+4\nu\nonumber\\
a_2 &=& -h-1-24\nu\nonumber\\
a_3 &=& -\frac32 (h+1)+6\nu\nonumber\\
a_4 &=& -20 \nu   \,.
\end{eqnarray}

A similar resummation of the 2PM-accurate $g_S$ is given by
\begin{eqnarray}
g_{S}^{\rm 2PM resum}&=&\frac{c_0+c_1 u}{h(h+1)+c_2\nu u \gamma}\nonumber\\
&& +\frac{1}{\gamma}\frac{-2\nu h +3\nu u(2h-1)}{h^2 (h+1)^2}\nonumber\\
&& +\frac{1}{\gamma+1} \frac{d_0+d_1u}{\gamma+1+d_2 u}\,,
\end{eqnarray}
where
\begin{eqnarray}
c_0 &=& \frac{4(h+1+\nu)}{h+1} \nonumber\\
c_1 &=& -\frac{10(2 h-5)\nu}{h(h+1)} \nonumber\\
c_2 &=& \frac{35(h+1)}{4(h+1+\nu)}  \nonumber\\
d_0 &=& -4 h\nu\nonumber\\
d_1 &=& 8\nu(-4+3h)\nonumber\\
d_2 &=& -1\,.
\end{eqnarray} 
One can easily complete these expressions by including the 3PM (i.e., $O(u^2)$) known contributions,
Eqs. \eqref{gSs3pm},  \eqref{gSs3pm}, in the above resummed expressions.

\section{Concluding remarks}

Using the notion of spin holonomy (i.e. a suitable projection of the spacetime spin rotation),  introduced, at the first post-Minkowskian order, in Ref. \cite{Bini:2017xzy}, we have computed the second-order post-Minkowskian (2PM) 
corrections to the spin-orbit coupling of a gravitationally interacting, comparable-mass, two-body system.
We transcribed our $O(G^2)$-accurate computation of the spin rotation during  two-body scattering into
the corresponding determination of the gyrogravitomagnetic ratios $g_S$ and $g_{S*}$ entering the effective one-body (EOB) Hamiltonian description of linear-in-spin coupling effects. Contrary to the previous, post-Newtonian-based, knowledge of $g_S$ and $g_{S*}$, our
2PM-accurate results are exact in $v/c$, and  represent a new direction for improving  the EOB conservative dynamics of 
a spinning two-body system. 

The computations have been performed in $x$-space by using the explicit form of the 2PM metric generated by two spinless bodies, available since 1981 \cite{Bel:1981be}. 
Most of our results have been double-checked by  Fourier-space ($k$-space) computations (see Appendix A).

We have indicated ways of resumming our final results so as to use them for defining new versions of 
spinning EOB codes. 
In view of future, forthcoming high signal-to-noise-ratio gravitational-wave observations, we think that 
the improved analytical knowledge of the conservative dynamics of a spinning two-body system brought by the present work
might play a useful role.
 
Similarly to the considerations recently made in \cite{Damour:2017zjx} for the orbital scattering, 
it would be interesting to explore whether one
can extract the classical spin scattering angle from quantum gravitational amplitudes. This would offer a new
avenue for further  improving the knowledge of spin-orbit couplings in gravitationally interacting systems.
Note that our classical 2PM-accurate result corresponds to the quantum one-loop level.

\appendix
\section{Fourier space computations}

Most of the terms computed above in $x$-space to obtain the spin holonomy matrix component $\Lambda_{(1)}{}^1{}_2$ can be validated by performing analogous calculations in Fourier space. The well-known Fourier-space results
\begin{eqnarray}
\int \frac{d^3 k}{(2\pi)^3}\frac{ e^{i\vec k\cdot \vec x}}{k^2}&=& +\frac{1}{4\pi |x|}\nonumber\\
\int \frac{d^2 K}{(2\pi)^2} \frac{e^{i\vec K\cdot \vec X}}{K^2}&=&-\frac{\ln |X|}{2\pi }\,,
\end{eqnarray}
where $\vec k, \vec x$ are vectors of the Euclidean 3-space and $\vec K,\vec X$ are vectors in the Euclidean 2-plane, will be often used 
in the following.

Let us recall that at the linear order in $G$ the gothic metric satisfies
\beq
\Box {\sf h}_{\alpha\beta}=+16 \pi G T_{\alpha\beta}\,.
\eeq
The source $T_{\alpha\beta}$ corresponds to the two particles $m_1$ and $m_2$ and
\beq
T_{\alpha\beta}(x)=T^{(m_1)}_{\alpha\beta}(x)+T^{(m_2)}_{\alpha\beta}(x)\,,
\eeq
where, for example
\beq
T^{(m_2)}_{\alpha\beta}(x)=m_2\int d\tau' u_2{}_\alpha (\tau')u_2{}_\beta (\tau') \delta^{(4)}(x-z_2{}(\tau'))\,.
\eeq
The part of the metric generated by $T^{(m_2)}$  turns out to be given by
\beq
{\sf h}_{\alpha\beta}(x)= -16 \pi G \int \frac{d^4 k}{(2\pi)^4}\frac{e^{i k \cdot x}}{k^2}T_{(m_2)}{}_{\alpha \beta }(k)\,,
\eeq
where
\begin{eqnarray}
T^{(m_2)}{}^{\alpha'\beta'}(k)&=&\int d^4 x e^{-ik\cdot x}T^{(m_2)}{}^{\alpha '\beta'}(x)\\
&=& m_2\int d\tau' u_2{}^\alpha (\tau')u_2{}^\beta (\tau') e^{-ik \cdot z_2(\tau')}\,.\nonumber
\end{eqnarray}

Therefore
\begin{eqnarray}
{\sf h}_{\alpha\beta}(x)&=& -16 \pi G m_2\int \frac{d^4 k}{(2\pi)^4}\frac{1}{k^2}\times \nonumber\\
&&  \int d\tau' u_2{}_\alpha (\tau')u_2{}_\beta (\tau') e^{ik \cdot (x- z_2(\tau'))}\,,
\end{eqnarray}
and
\beq
{\sf h}(x)=  16 \pi G m_2\int \frac{d^4 k}{(2\pi)^4}\frac{1}{k^2} \int d\tau'   e^{ik \cdot (x- z_2(\tau'))}\,,
\eeq
since $u_2(\tau')\cdot u_2(\tau')=-1+O(G^2)$.\\

\noindent We will compute the following tensors
\begin{eqnarray}
\label{B_e_C_defs}
{\mathcal B}_{\mu\alpha} &=&\int \partial_\mu {\sf h}_{\alpha\beta}(x) dz^\beta= \int d\tau \left(\partial_\mu {\sf h}_{\alpha\beta}(x)\right)_{x=z_1(\tau)} u_1^\beta(\tau)\nonumber\\
& =&  -16 \pi G  m_2 \int d\tau  u_1^\beta(\tau) \times \nonumber\\
&& \int \frac{d^4 k}{(2\pi)^4}\frac{ik_\mu }{k^2} \int d\tau' u_2{}_\alpha (\tau')u_2{}_\beta (\tau') e^{ik \cdot (z_1(\tau)- z_2(\tau'))}\nonumber\\
{\mathcal C}_{\mu}{}^{\beta} &=&\int \partial_\mu {\sf h}(x) dz^\beta= \int d\tau \left(\partial_\mu {\sf h}\right)_{x=z_1(\tau)} u_1^\beta(\tau) \nonumber\\
&=&  16 \pi G m_2\int  u_1^\beta(\tau)  d\tau \times \nonumber\\
&&  \int \frac{d^4 k}{(2\pi)^4}\frac{ik_\mu}{k^2} \int d\tau'   e^{ik \cdot (z_1(\tau)- z_2(\tau'))}\,.
\end{eqnarray}
They enter $\Lambda_{(1)}{}^1{}_2$ through the relation
\begin{eqnarray}
\label{eqA10}
\Lambda_{(1)}{}^1{}_2&=&\frac12 \left({\mathcal B}_{ xy }-{\mathcal B}_{yx}  \right)-\frac14 \left({\mathcal C}_{ xy }-{\mathcal C}_{yx} \right)\nonumber\\
&=&{\mathcal B}_{[xy]}-\frac12 {\mathcal C}_{[xy]}\nonumber\\
&\equiv & \Lambda_{{\mathcal B}}{}^1{}_2+\Lambda_{{\mathcal C}}{}^1{}_2\,,
\end{eqnarray}
where one has used the standard notation for antisymmetrization of indices. In fact, from the first-order relation 
\begin{eqnarray}
2 \Gamma^x{}_{y\mu}u_1^\mu &=&
\partial_x {\sf h}_{y\gamma}u_1^\gamma -\partial_y {\sf h}_{x\gamma}u_1^\gamma\nonumber\\
&& -\frac12 u_1^y \partial_x {\sf h}+\frac12 u_1^x \partial_y {\sf h}\,,
\end{eqnarray}
one immediately obtains Eq. \eqref{eqA10}.
  
\subsection{The case of straight lines}

In this limit the worldlines of the two particles read
\beq
z_1(\tau)=b_0\partial_x +u_1^-\tau \,,\qquad z_2(\tau')=u_2^-\tau'
\eeq
and $b_0=b$ coincides with the impact parameter.

\begin{enumerate}
  \item Computation of ${\mathcal B}_{\mu\alpha}$\\

\noindent 
Let us start analyzing the case of straight lines
\begin{eqnarray}
{\mathcal B}_{\mu\alpha}&=&  +16 \pi G m_2 \gamma   u_2^-{}_\alpha    \int \frac{d^4 k}{(2\pi)^4}\frac{ik_\mu }{k^2}e^{i k \cdot b_0}\times \nonumber\\
&&  \int d\tau  e^{i(k \cdot u_1^-)\tau}  \int d\tau'  
  e^{-i(k \cdot u_2^-)\tau'}
\nonumber\\
&=&  +16 \pi G m_2 \gamma   u_2^-{}_\alpha \times \nonumber\\
&& \frac{\partial}{\partial b_0^\mu}   \int \frac{d^4 k}{(2\pi)^2}\frac{e^{i k \cdot b_0}}{k^2} \delta (k \cdot u_1^-)  
\delta (k \cdot u_2^-)\,.
\end{eqnarray}
Expressing $\delta (k \cdot u_1^-)\delta (k \cdot u_2^-)$ in the usual coordinate system we find
\beq
\delta (k \cdot u_1^-)  \delta (k \cdot u_2^-)=\frac{1}{\sqrt{\gamma^2-1}}\delta(k^0)\delta(k^y)\,,
\eeq
and then, recalling that $\frac{\partial b_0}{\partial b_0^\mu}=\frac{b_0{}_\mu}{b_0}$, we find
\begin{eqnarray}
{\mathcal B}_{\mu\alpha}&=&16 \pi G m_2 \frac{\gamma}{\sqrt{\gamma^2-1}}   u_2^-{}_\alpha \frac{\partial}{\partial b_0^\mu}   \int \frac{dk^x dk^z}{(2\pi)^2}\frac{e^{i k \cdot b_0}}{k^2}\nonumber\\
&=&-16 \pi G m_2 \frac{\gamma}{\sqrt{\gamma^2-1}}   u_2{}^-_\alpha \frac{\partial}{\partial b_0^\mu} \frac{1}{2\pi} \ln b_0 \nonumber\\
&=& -8 G m_2 \frac{\gamma}{\sqrt{\gamma^2-1}}   \frac{b_0{}_\mu}{b_0^2}u_2^-{}_\alpha \,,
\end{eqnarray}
implying
\beq
{\mathcal B}_{xy}=8 G m_2\frac{\gamma }{b_0}\,,\qquad\qquad  {\mathcal B}_{yx}=0\,.
\eeq

\item Computation of ${\mathcal C}_{\mu\alpha}$\\

\noindent  
In this case we have
  \begin{eqnarray}
{\mathcal C}_{\mu\beta}
&=& -16 \pi G m_2 \delta _\beta^0 \int  \frac{d^4 k}{(2\pi)^4}\frac{ik_\mu }{k^2}e^{i k \cdot b_0} \times \nonumber\\
&&  \int d\tau  e^{ik \cdot u_1^- \tau}  \int d\tau'  e^{-ik \cdot u_2^- \tau'}\nonumber\\
&=& -16 \pi G m_2 \delta _\beta^0 \frac{\partial}{\partial b_0^\mu } \int  \frac{d^4 k}{(2\pi)^4}\nonumber\\
&& \frac{ e^{i k \cdot b_0} }{k^2}(2\pi)^2\delta (k\cdot u_1^-)\delta (k\cdot u_2^-)
\nonumber\\
&=&  -16 \pi G m_2 \delta _\beta^0 \frac{1}{\sqrt{\gamma^2-1}} \frac{b_0^\mu }{b_0^2}\,,
\end{eqnarray} 
so that ${\mathcal C}_{xy}={\mathcal C}_{yx}=0$.

\end{enumerate}

\noindent
The final result for $\Lambda_{(1)}^1{}_2$ at the 1PM order coincides with the well-known result
\beq
\Lambda_{(1)}^1{}_2=-\frac12 {\mathcal B}_{xy}=4 G m_2 \frac{ \gamma}{b_0}\,.
\eeq

In this special case (straight lines) one can easily compute $\partial_\alpha {\sf h}_{\beta \mu}(x)$, namely
\beq
\partial_\alpha {\sf h}_{\beta \mu}(x)= C_\alpha u'_-{}_\beta  u'_-{}_\mu\,,\qquad \partial_\alpha {\sf h}(x)=- C_\alpha
\eeq 
where
\begin{eqnarray}
C_\alpha(x) &=& -16 \pi G m_2 \int \frac{d^4k}{(2\pi)^4}\frac{ik_\alpha}{k^2} (2\pi) e^{ik\cdot (x-z_2(0))}\delta (k\cdot u_2^-)\nonumber\\
&=& -16 \pi G m_2 \frac{\partial}{\partial \xi^\alpha}\int \frac{d^4k}{(2\pi)^3}\frac{e^{ik\cdot \xi }}{k^2}  \delta (k^0+\frac{\sqrt{\gamma^2-1}}{\gamma}k^y)\nonumber\\
&=& -16 \pi G m_2 \frac{\partial}{\partial \xi^\alpha}\frac{1}{4\pi}\frac{1}{|\xi|}=4Gm_2 \frac{\xi_\alpha}{|\xi|^3}\,,
\end{eqnarray}
where
\begin{eqnarray}
\xi &=& P(u_2^-)x=x+(u_2^-\cdot x )u_2^-\nonumber\\
&=& x\partial_x +(\gamma y +\sqrt{\gamma^2-1}\, t)e(u_2^-)_2 +z\partial_z\,, 
\end{eqnarray}
with 
\beq
 e(u_2^-)_2= -\sqrt{\gamma^2-1}\partial_t +\gamma \partial_y\,.
\eeq
Along the particle $m_1$ worldline
\beq
\xi \to  b_0\partial_x +t\,  \sqrt{\gamma^2-1} e(u_2^-)_2 \,,
\eeq
so that
\begin{eqnarray}
\frac{\xi_t}{|\xi|^3}&=&\frac{(\gamma^2-1)t}{D(t)^3}\nonumber\\ 
\frac{\xi_x}{|\xi|^3}&=&\frac{b_0}{D(t)^3}\nonumber\\
\frac{\xi_y}{|\xi|^3}&=&\frac{\gamma \sqrt{\gamma^2-1}\,t }{D(t)^3}\,.
\end{eqnarray}
Replacing
\beq
\partial_\alpha {\sf h}_{\beta \mu}(x)= 4Gm_2 \frac{\xi_\alpha}{|\xi|^3} u'_-{}_\beta  u'_-{}_\mu
=C_\alpha u'_-{}_\beta  u'_-{}_\mu\,,
\eeq
implying 
\beq
u'_- \cdot C =0\,,\qquad u'_-{}^\alpha \partial_\alpha {\sf h}_{\beta \mu}(x)=0\,,
\eeq
in the connection term (also evaluated along the worldline of the particle $m_1$, with $x=u_1^- \tau +b_0$ and $C^\alpha=C^\alpha(\tau)$), namely

\begin{eqnarray}
\omega_1(\tau)^\alpha{}_\beta &=&\frac12 \left( \partial^\alpha {\sf h}_{\beta 0}-\partial_\beta {\sf h}^\alpha{}_{0}-\partial_0 {\sf h}^\alpha{}_\beta \right)\nonumber\\
&&+\frac14  (\delta^\alpha_0 \partial_\beta {\sf h}+\delta^\alpha_\beta \partial_0 {\sf h}+\delta^0_\beta \partial^\alpha  {\sf h})\bigg|_{x=x(\tau)}
\end{eqnarray}
we find
\begin{eqnarray}
\omega_1(\tau)^\alpha{}_\beta &=&\left\{ \frac12 \left[-\gamma \left(C^\alpha u_2^-{}_\beta -u_2^-{}^\alpha C_\beta \right) \right.\right. \nonumber\\
&&\left. -C_0 u_2^-{}^\alpha u_2^-{}_\beta\right]\nonumber\\
&& \left.-\frac14 [\delta^\alpha_0C_\beta +\delta^\alpha_\beta C_0+\delta^0_\beta C^\alpha]\right\}d\tau\nonumber\\
&=& -\frac{\gamma}{2}[C\wedge u_2^-]^\alpha{}_\beta +\frac14 [C\wedge u_1^- ]^\alpha{}_\beta \nonumber\\
&& -\frac12 C_0 \left(u_2^-{}^\alpha u_2^-{}_\beta+\frac12 \delta^\alpha_\beta\right)\nonumber\\
&=&[ {\sf C}_1(\tau)^\alpha{}_\beta +{\sf C}_2(\tau)^\alpha{}_\beta ]d\tau\,,
\end{eqnarray}
or explicitly
\begin{widetext}
\beq
\omega_1(\tau)^\alpha{}_\beta  =
\begin{pmatrix}
\frac{C^0(1-2\gamma^2)}{4} & -\frac{C^1(1-2\gamma^2)}{4} & \frac{C^2(1-2\gamma^2)}{4} & 0 \cr
-\frac{C^1(1-2\gamma^2)}{4} & \frac{C^0}{4} & \frac{\gamma \sqrt{\gamma^2-1}C^1}{2} & 0 \cr
-\frac{C^2(1-2\gamma^2)}{4}+\gamma \sqrt{\gamma^2-1}C^0 & -\frac{\gamma \sqrt{\gamma^2-1}C^1}{2} & -\frac{C^0(1-2\gamma^2)}{4} &0 \cr
0 & 0 & 0 & \frac{C^0}{4}\,.
\end{pmatrix}
\eeq
\end{widetext}
We can compute then
\begin{eqnarray}
\omega_1(\tau)^\alpha{}_\beta  \omega_1(\tau')^\beta{}_\mu &=& [{\sf C}_1(\tau)^\alpha{}_\beta +{\sf C}_2(\tau)^\alpha{}_\beta ]\times \\
&& [{\sf C}_1(\tau')^\beta{}_\mu +{\sf C}_2(\tau')^\beta{}_\mu ] d\tau d\tau'\,,\nonumber
\end{eqnarray}
and finally, using the abbreviated notation $C^\alpha=C^\alpha(\tau)$ and $C'{}^\alpha =C'{}^\alpha(\tau')$, 
\begin{eqnarray}
\omega_1(\tau)^x{}_\beta  \omega_1(\tau')^\beta{}_y &=&\left\{\frac{2\gamma^2-1}{8}\left[\frac12 (2\gamma^2-1)C'{}^2\right.\right.\nonumber\\
&+& \left.\gamma \sqrt{\gamma^2-1}C'{}^0 \right]C^1\nonumber\\
&+& \left.
\frac{\gamma\sqrt{\gamma^2-1}}{8}C^0 C'{}^1\right\}  d\tau' d\tau\,.
 \end{eqnarray}
Integrating over $\tau'\in (-\infty, \tau)$ and then over $\tau\in (-\infty, +\infty)$ gives
\beq
\int_{-\infty}^\infty \int_{-\infty}^\tau \omega_1(\tau)^x{}_\beta  \omega_1(\tau')^\beta{}_y = -\frac{\pi G^2 m_2{}^2}{2}\frac{\gamma (4\gamma^2-3)}{b_0^2 (\gamma^2-1)}
\eeq
which coincides with the previously obtained result.

\section{The general case: computation of ${\mathcal B}_{\mu\alpha}$}

Start from Eq. \eqref{B_e_C_defs} and insert the four velocities
\beq
u_1^\beta(\tau)=u_1^-{}^\beta +\delta u_1^\beta (\tau)\,,\quad
u_2{}^\beta(\tau')=u_2^-{}^\beta +\delta u_2{}^\beta (\tau')
\eeq
and worldlines
\begin{eqnarray}
z_1^\beta(\tau) &=& z_1(0)^\beta +u_1^-{}^\beta \tau  +\delta z_1^\beta (\tau)\,,\nonumber\\
z_2{}^\beta(\tau')&=& z_2(0)^\beta +u_2^-{}^\beta \tau'  +\delta z_2{}^\beta (\tau')\,,
\end{eqnarray}
with   $b_0^\beta=z_1(0)^\beta$ and $z_2(0)^\beta$ vanishing, namely
\begin{eqnarray}
\label{gen_cal_B2}
{\mathcal B}_{\mu\alpha}=  -16 \pi G m_2  \int \frac{d^4 k}{(2\pi)^4}\frac{ik_\mu }{k^2} e^{ik \cdot b_0} F_\alpha (k)
\,,
\end{eqnarray}
where
\begin{widetext}
\begin{eqnarray}
F^\alpha(k)&=&\iint  d\tau  d\tau'  (u_1^-{}^\beta+\delta u_1^\beta(\tau)) (u_2^-{}_\beta+\delta u_2{}_\beta (\tau')) (u_2^-{}^\alpha +\delta u_2{}^\alpha (\tau')) e^{ik \cdot (u_1^-\tau +\delta z_1(\tau)-u_2^- \tau' -\delta z_2(\tau')}\nonumber\\
&=& \iint e^{ik \cdot  u_1^-\tau} d\tau  e^{-ik \cdot u_2^- \tau'}     d\tau' \{u_2^-{}^\alpha [-\gamma +u_1^-\cdot \delta u_2 +u_2^-\cdot \delta u_1 -i\gamma k \cdot  \delta z_1(\tau)+i  \gamma k \cdot \delta z_2(\tau')]-\gamma \delta u_2{}^\alpha (\tau') \}\,.   
\end{eqnarray}
\end{widetext}

The six terms entering $F^\alpha(k)$ can be re-written as
\begin{eqnarray}
F^\alpha(k)
&=&-3\gamma  u_2^-{}^\alpha  (2\pi)^2 \delta(k \cdot  u_1^-)\delta(k \cdot u_2^-)    \nonumber\\ 
&+& (2\pi)  \delta(k \cdot  u_1^-) u_2^-{}^\alpha u_1^-\cdot \int  e^{-ik \cdot u_2^- \tau'}     d\tau'    \delta u_2(\tau') \nonumber\\
&+& (2\pi) \delta (k\cdot u_2^-) u_2^-{}^\alpha u_2^- \cdot \int e^{ik \cdot  u_1^-\tau} d\tau    \delta u_1 (\tau) \nonumber\\
\end{eqnarray}
\begin{eqnarray}
&+& (2\pi) \delta (k\cdot u_2^-) \gamma u_2^-{}^\alpha   \int e^{ik \cdot  u_1^-\tau-i k \cdot \delta z_1(\tau)} d\tau   \nonumber\\
&+& (2\pi)  \delta(k \cdot  u_1^-)  \gamma u_2^-{}^\alpha  \int    e^{-ik \cdot u_2^- \tau'+i k \cdot  \delta z_2(\tau')}     d\tau'  \nonumber\\  
&-& (2\pi)  \delta(k \cdot  u_1^-) \gamma \int   e^{-ik \cdot u_2^- \tau'}     d\tau'  \delta u_2{}^\alpha (\tau')  \nonumber\\
&\equiv & \sum_{i=1}^6 F_i^\alpha(k)\,,
\end{eqnarray}
and using Eq.  \eqref{gen_cal_B2} one finds the corresponding contributions to ${\mathcal B}_{\mu\alpha}$, 
\begin{eqnarray}
\label{gen_cal_B2_repeated}
{\mathcal B}_{\mu\alpha}&\equiv & \sum_{n=1}^6 {\mathcal B}^{(1)}_{\mu\alpha}\\
&=& -16 \pi G  m_2  \int \frac{d^4 k}{(2\pi)^4}\frac{ik_\mu e^{ik \cdot b_0}  }{k^2}\sum_{n=1}^6 F_{n\alpha} (k)\,,\nonumber
\end{eqnarray}
identifying the various terms contributing to ${\mathcal B}_{\mu\alpha}$.

\subsection{Computation of ${\mathcal B}^{(1)}_{\mu\alpha}$}
In this case  
\begin{eqnarray}
\label{gen_cal_B2}
{\mathcal B}^{(1)}_{\mu\alpha}&=& 48\gamma  u_2^-{}_\alpha  \pi G m_2  \frac{\partial}{\partial b_0^\mu }\times \nonumber\\
&&
\int \frac{d^4 k}{(2\pi)^2}\frac{ e^{ik \cdot b_0}  }{k^2} \delta(k \cdot  u_1^-)\delta(k \cdot u_2^-)\nonumber\\
&=& 48\gamma  u_2^-{}_\alpha  \pi G m_2 \frac{1}{\sqrt{\gamma^2-1}}\frac{\partial}{\partial b_0^\mu }\int \frac{dk^x dk^z}{(2\pi)^2}\frac{ e^{ik_x b_0}  }{k_x^2+k_z^2} \nonumber\\
&=& -24  u_2^-{}_\alpha   G m_2 \frac{\gamma}{\sqrt{\gamma^2-1}}\frac{\partial}{\partial b_0^\mu }\ln b_0 \,,
\end{eqnarray}
so that, using the relation $\frac{\partial}{\partial b_0^\mu } b_0=\frac{b_0{}_\mu}{b_0}$, 
\beq
{\mathcal B}^{(1)}_{\mu\alpha}=-24 G m_2   \frac{\gamma}{\sqrt{\gamma^2-1}}\frac{ b_0{}_\mu }{b_0^2 }u_2^-{}_\alpha \,.
\eeq
In particular
\beq
{\mathcal B}^{(1)}_{xy}=24 G m_2    \frac{\gamma}{b_0}\,,\qquad\qquad  {\mathcal B}^{(1)}_{yx}=0 \,.
\eeq

\subsection{Computation of ${\mathcal B}^{(2)}_{\mu\alpha}$}

In this case
\begin{eqnarray}
{\mathcal B}^{(2)}_{\mu\alpha}
&=& -16 \pi G m_2 u_2^-{}_\alpha u_1^-\cdot  \int d\tau'    \delta u_2(\tau')  \nonumber\\
&&  \int \frac{d^4 k}{(2\pi)^3}\frac{ik_\mu e^{ik \cdot (b_0-u_2^- \tau')}  }{k^2}  \delta(k^0) \,.    \end{eqnarray}
Introducing the vectors
\beq
\xi=b_0-u_2^- \tau'\,,\qquad  \eta =P(u_1^-) \xi=b_0+\sqrt{\gamma^2-1}\tau' \partial_y\,.
\eeq
and proceeding as before one easily finds then
\beq
{\mathcal B}^{(2)}_{xy}=
-2\pi  G^2 m_1 m_2  \frac{\gamma (2\gamma^2-3) }{b_0^2}\,,\qquad  {\mathcal B}^{(2)}_{yx}=0\,.
\eeq

\subsection{Computation of ${\mathcal B}^{(3)}_{\mu\alpha}$}

In this case
\begin{eqnarray}
{\mathcal B}^{(3)}_{\mu\alpha}&=& -16 \pi G m_2  u_2^-{}_\alpha u_2^-\cdot \int  d\tau    \delta u_1 (\tau) \nonumber\\
&&  \int \frac{d^4 k}{(2\pi)^3}\frac{ik_\mu e^{ik \cdot (b_0+u_1^-\tau)}  }{k^2} \delta (k\cdot u_2^-)  \nonumber\\
&=& -16 \pi G m_2  u_2^-{}_\alpha u_2^-\cdot \int  d\tau    \delta u_1 (\tau)  \nonumber\\
&& \int \frac{d^4 k}{(2\pi)^3}\frac{i[P(u_2^-)k]_\mu e^{iP(u_2^-)k \cdot (b_0+u_1^-\tau)}  }{|P(u_2^-)k|^2} \nonumber\\
&& \delta (\gamma k^0+\sqrt{\gamma^2-1}k^y)  \,,
\end{eqnarray}
where we have replaced $k$ by $P(u_2^-)k$ because of the condition $u_2\cdot k=0$.
The various quantities involved here reduce to
\begin{eqnarray}
P(u_2^-)k  
&=& k^x \partial_x+(\gamma^{-1}k^y)e(u_2^-)_2+k^z\partial_z\nonumber\\
|P(u_2^-)k|^2&=& (k^x)^2+(\gamma^{-1}k^y)^2+(k^z)^2\nonumber\\
P(u_2^-)k \cdot (b_0+u_1^-\tau)&=& k^x b_0 +\frac{\sqrt{\gamma^2-1}}{\gamma} k^y \tau\,,
\end{eqnarray}
and defining
\beq
\xi = b_0+u_1^-\tau\,,\qquad \eta=P(u_2^-)\xi\,,
\eeq
we find then 
\beq
{\mathcal B}^{(3)}_{xy}= -2\pi  G^2 m'{}^2  \frac{\gamma  (2\gamma^2-3)}{b_0^2}\,,\qquad  {\mathcal B}^{(3)}_{yx}=0\,.
\eeq

\subsection{Computation of ${\mathcal B}^{(4)}_{\mu\alpha}$}

In this case we introduce the   vector
\begin{eqnarray}
\xi &=& b_0 + u_1^-\tau -\delta z_1(\tau)\,,\nonumber\\
{}[P(u_2^-) k] \cdot \xi &=& k^x (b_0 -\delta z_1^x) \nonumber\\
&+& (\gamma^{-1} k^y) [ \sqrt{\gamma^2-1}  \tau -  \gamma \delta z_1^y]\,,
\end{eqnarray}
with
\beq
P(u_2^-)k=k^x\partial_x +(\gamma^{-1}k^y)e(u_2^-)_2 +k^z \partial_z\,.
\eeq
Its projection orthogonal to $u_2^-$ is the four-vector
\beq
\eta = P(u_2^-)\xi 
\eeq
(not to be confused with the Minkowski metric) which reads
\begin{eqnarray}
\eta &\equiv & P(u_2^-)\xi \nonumber\\
&=& (b_0-\delta z_1^x) \partial_x +[\sqrt{\gamma^2-1}\tau -\gamma \delta z_1^y](\gamma \partial_y -\sqrt{\gamma^2-1}\partial_t)\nonumber\\
&=& (b_0-\delta z_1^x) \partial_x +[\sqrt{\gamma^2-1}\tau -\gamma \delta z_1^y]e(u_2^-)_2 \nonumber\\
&=& \eta^x \partial_x +\eta^y e(u_2^-)_2 \,.
\end{eqnarray}
Therefore
\begin{widetext}

\begin{eqnarray}
{\mathcal B}^{(4)}_{\mu\alpha}
&=& -16 \pi G m_2 \gamma u_2^-{}_\alpha   \int d\tau\int \frac{d^4 k}{(2\pi)^3}\frac{ik_\mu e^{ik \cdot b_0}  }{k^2}  \delta (k\cdot u_2^-)  e^{ik \cdot ( u_1^-\tau- \delta z_1(\tau))}  \nonumber\\
&=& -16 \pi G m_2 \gamma u_2^-{}_\alpha   \int d\tau\int \frac{d^4 k}{(2\pi)^3}\frac{i[P(u_2^-)k]_\mu e^{iP(u_2^-)k \cdot (b_0 + u_1^-\tau -\delta z_1(\tau)}  }{|P(u_2^-)k|^2}  \delta (k\cdot u_2^-)    \nonumber\\
&=& 4  G m_2 \gamma u_2^-{}_\alpha   \int d\tau  \frac{\eta_\mu}{[(b_0-\delta z_1^x)^2+(\sqrt{\gamma^2-1}\tau -\gamma  \delta z_1^y)^2]^{3/2}}\,,
\end{eqnarray}
where, we recall
\beq
\eta_x = b_0-\delta z_1^x  \,,\qquad \eta_y= \sqrt{\gamma^2-1}\tau -\gamma \delta z_1^y\,.
\eeq
In particular
\beq
{\mathcal B}^{(4)}_{xy}= -8 G m' \frac{\gamma}{b_0}  -4 \frac{\gamma}{ b_0^2}   m'{}^2 G^2\left(-\pi\frac{2\gamma^2+1}{2} +2\frac{ 2\gamma^2-1 }{ \gamma^2-1 } \right)
\,,\qquad  {\mathcal B}^{(4)}_{yx}=0\,.
\eeq

\subsection{Computation of ${\mathcal B}^{(5)}_{\mu\alpha}$}
Let us introduce the   vector
\beq
\xi = b_0 -u_2^-\tau' +\delta z_2(\tau')\,,\qquad [P(u_1^-) k] \cdot \xi =k^x (b_0+\delta z_2{}^x)+k^y (\sqrt{\gamma^2-1}\tau' +\delta z_2{}^y)
\eeq
such that  
\beq
\eta = P(u_1^-)\xi =(b_0+\delta z_2{}^x)\partial_x +(\sqrt{\gamma^2-1}\tau' +\delta z_2{}^y)\partial_y\,.
\eeq
We find
\begin{eqnarray}
{\mathcal B}^{(5)}_{\mu\alpha}&=& -16 \pi G m_2 \gamma u_2^-{}_\alpha   \int  d\tau' \int \frac{d^3 k}{(2\pi)^3}\frac{i(P(u_1^-)k)_\mu e^{i\vec k \cdot ( b_0-P(u_1^-)u_2^- \tau'+P(u_-)\delta z_2(\tau'))}  }{\vec k^2}      \nonumber\\
&=& -4 G m_2\gamma u_2^-{}_\alpha   \int  d\tau' \frac{\partial}{\partial  \eta^\mu }\frac{1}{| b_0-P(u_1^-)u_2^- \tau'+P(u_1^-)\delta z_2(\tau'))|} \nonumber\\
&=&  4 G m_2\gamma u_2^-{}_\alpha   \int  d\tau'  \frac{ \eta_\mu}{[(b_0+\delta z_2(\tau')^x)^2+(\sqrt{\gamma^2-1}\tau' +\delta z_2(\tau')^y)^2]^{3/2}}  \,.
 \end{eqnarray}
In particular
\beq
{\mathcal B}^{(5)}_{xy}= -8 G m_2  \frac{\gamma   }{b_0}    -4 G^2 m_1 m_2  \left(  -\frac12 (2\gamma^2+1)\pi  +2\frac{ 2\gamma^2-1 }{ \gamma^2-1 }\right) \frac{\gamma   }{b_0^2} \,,\qquad {\mathcal B}^{(5)}_{yx}=0 \,.
\eeq

\subsection{Computation of ${\mathcal B}^{(6)}_{\mu\alpha}$}

Let us introduce the vectors
\beq
\xi = b_0 -u_2^-\tau'\,,\qquad \eta=P(u_1^-)\xi=b_0+\sqrt{\gamma^2-1}\tau' \partial_y\,.
\eeq
We have then
\begin{eqnarray}
{\mathcal B}^{(6)}_{\mu\alpha}&=& +16 \pi G m_2\gamma  \int  d\tau'\delta u_2 (\tau'){}_\alpha  \int \frac{d^4 k}{(2\pi)^3}\frac{ik_\mu e^{ik \cdot b_0}  }{k^2}   \delta(k \cdot  u_1^-) e^{-ik \cdot u_2^- \tau'}       \nonumber\\
&=& +16 \pi G m_2  \gamma     \int d\tau'   \delta u_2(\tau'){}_\alpha  \frac{\partial}{\partial \eta^\mu }\int \frac{d^3 k}{(2\pi)^3}\frac{ e^{i\vec k \cdot (b_0-P(u_1^-)u_2^- \tau')}  }{\vec k^2}  \nonumber\\
&=& +4 G m_2  \gamma  \int d\tau' \delta u_2(\tau'){}_\alpha   \frac{-\eta_\mu}{D(\tau')^3}  \,.
\end{eqnarray}
In particular
\beq
{\mathcal B}^{(6)}_{xy}=  -2  G^2 m_1m_2  \pi   \frac{\gamma^2 (2\gamma^2-3)}{ (\gamma^2-1)}\frac{\gamma}{b_0^2} \,,\qquad
{\mathcal B}^{(6)}_{yx}= -2  G^2 m_1m_2  \pi  \frac{(2\gamma^2-1)}{\gamma^2-1}\frac{\gamma}{b_0^2}
\,.
\eeq

\section{The general case: computation of ${\mathcal C}_{\mu\alpha}$}

Let us recall
\begin{eqnarray}
{\mathcal C}_{\mu\beta}  
&=&  16 \pi G m_2\int  u_1{}_\beta(\tau)  d\tau  \int \frac{d^4 k}{(2\pi)^4}\frac{ik_\mu}{k^2} \int d\tau'   e^{ik \cdot b_0 }e^{i(u_1^-\tau+\delta z_1(\tau) - u_2^-\tau'-\delta z_2(\tau'))}\nonumber\\
&=& 16 \pi G m_2    \iint   d\tau d\tau' \int \frac{d^4 k}{(2\pi)^4} \frac{ik_\mu e^{ik \cdot b_0 }}{k^2}  e^{ik \cdot u_1^-\tau}
e^{ -ik \cdot  u_2^-\tau'} [u_1^-{}_\beta \left(e^{ik \cdot  \delta z_1(\tau) } + e^{ -ik \cdot  \delta z_2(\tau')}-1\right)+\delta u_1{}_\beta(\tau) ]
\end{eqnarray}
We will compute separately  the following four terms
\begin{eqnarray}
{\mathcal C}_{\mu\beta}^{(1)}  
&=& 16 \pi G m_2  u_1^-{}_\beta    \iint   d\tau d\tau' \int \frac{d^4 k}{(2\pi)^4} \frac{ik_\mu e^{ik \cdot b_0 }}{k^2}  e^{ik \cdot u_1^-\tau}
e^{ -ik \cdot  u_2^-\tau'}   e^{ik \cdot  \delta z_1(\tau) } \nonumber\\
{\mathcal C}_{\mu\beta}^{(2)}  &=& 16 \pi G m_2  u_1^-{}_\beta   \iint   d\tau d\tau' \int \frac{d^4 k}{(2\pi)^4} \frac{ik_\mu e^{ik \cdot b_0 }}{k^2}  e^{ik \cdot u_1^-\tau}
e^{ -ik \cdot  u_2^-\tau'}   e^{ -ik \cdot  \delta z_2(\tau')} 
\nonumber\\
{\mathcal C}_{\mu\beta}^{(3)}  &=& -16 \pi G m_2   u_1^-{}_\beta   \iint   d\tau d\tau' \int \frac{d^4 k}{(2\pi)^4} \frac{ik_\mu e^{ik \cdot b_0 }}{k^2}  e^{ik \cdot u_1^-\tau}
e^{ -ik \cdot  u_2^-\tau'}  \nonumber\\
{\mathcal C}_{\mu\beta}^{(4)}  &=& 16 \pi G m_2    \iint   d\tau d\tau' \int \frac{d^4 k}{(2\pi)^4} \frac{ik_\mu e^{ik \cdot b_0 }}{k^2}  e^{ik \cdot u_1^-\tau}
e^{ -ik \cdot  u_2^-\tau'}  \delta u_1{}_\beta(\tau) \,.
\end{eqnarray}

\subsection{Computation of ${\mathcal C}^{(1)}_{\mu\alpha}$}

We find
\begin{eqnarray}
{\mathcal C}_{\mu\beta}^{(1)}  
&=& 16 \pi G m_2  u_1^-{}_\beta    \iint   d\tau d\tau' \int \frac{d^4 k}{(2\pi)^4} \frac{ik_\mu e^{ik \cdot b_0 }}{k^2}  e^{ik \cdot u_1^-\tau}
e^{ -ik \cdot  u_2^-\tau'}   e^{ik \cdot  \delta z_1(\tau) } \nonumber\\
&=& 16 \pi G m_2  u_1^-{}_\beta   \int   d\tau  \int \frac{d^4 k}{(2\pi)^3} \frac{i k_\mu e^{ik \cdot b_0 }}{k^2}  e^{ik \cdot u_1^-\tau}
 e^{ik \cdot  \delta z_1(\tau) } \delta (k \cdot  u_2^-)  \nonumber\\
&=& 16 \pi G m_2   u_1^-{}_\beta    \int   d\tau  \int \frac{dk^x d(\gamma^{-1})k^y dk^z}{(2\pi)^3} \frac{i[P(u_2^-)k]_\mu e^{iP(u_2^-)k \cdot (b_0+u_1^-\tau +\delta z_1(\tau))}}{|P(u_2^-)k|^2}   
\end{eqnarray}
Let us introduce the vectors
\beq
 \xi =b_0+u_1^-\tau +\delta z_1(\tau)\,,\qquad  \eta =P(u_2^-) \xi\,,
\eeq
such that
\begin{eqnarray}
P(u_2^-)k\cdot  \xi &=& k^x (b_0+\delta z_1^x)+ (\gamma^{-1}k^y)[\sqrt{\gamma^2-1}\tau +\gamma \delta z_1^y]\nonumber\\
\eta &=& (b_0+\delta z_1^x)\partial_x +[\sqrt{\gamma^2-1}\tau +\gamma \delta z_1^y] e(u_2^-)_2\,.
\end{eqnarray}
A straightforward computation shows that
\beq
{\mathcal C}_{xy}^{(1)}={\mathcal C}_{yx}^{(1)}= 0\,.
\eeq

\subsection{Computation of ${\mathcal C}^{(2)}_{\mu\alpha}$}

\begin{eqnarray}
{\mathcal C}_{\mu\beta}^{(2)}  &=& 16 \pi G m_2  u_1^-{}_\beta    \iint   d\tau d\tau' \int \frac{d^4 k}{(2\pi)^4} \frac{ik_\mu e^{ik \cdot b_0 }}{k^2}  e^{ik \cdot u_1^-\tau}
e^{ -ik \cdot  u_2^-\tau'}   e^{ -ik \cdot  \delta z_2(\tau')} 
\nonumber\\
&=& 16 \pi G m_2  u_1^-{}_\beta     \int   d\tau' \int \frac{d^4 k}{(2\pi)^3} \frac{ik_\mu e^{ik \cdot b_0 }}{k^2} \delta(k \cdot u_1^-)  
e^{ -ik \cdot  u_2^-\tau'}   e^{ -ik \cdot  \delta z_2(\tau')}  \nonumber\\
&=& 16 \pi G m_2  u_1^-{}_\beta     \int   d\tau' \int \frac{d^3 k}{(2\pi)^3} \frac{i[P(u_1^-)k]_\mu e^{iP(u_1^-) k \cdot (b_0 -u_2^-\tau'- \delta z_2(\tau'))}}{|P(u_1^-) k|^2} \,.
\end{eqnarray}
Let us introduce the vectors
\beq
\xi  =b_0 -u_2^- -\tau'- \delta z_2(\tau')\,,\qquad \eta =P(u_1^-)\xi 
\eeq 
such that
\begin{eqnarray}
P(u_1^-)k \cdot \xi  &=& k^x (b_0-\delta z_2{}^x)+k^y[\sqrt{\gamma^2-1}\tau'-\delta z_2{}^y] \nonumber\\
\eta  &=&(b_0-\delta z_2{}^x)\partial_x+[\sqrt{\gamma^2-1}\tau'-\delta z_2{}^y]\partial_y \,.
\end{eqnarray}
Proceeding as before one finds
\beq
{\mathcal C}_{xy}^{(2)}={\mathcal C}_{yx}^{(2)}=0\,.
\eeq
 
\subsection{Computation of ${\mathcal C}^{(3)}_{\mu\alpha}$}

\begin{eqnarray}
{\mathcal C}_{\mu\beta}^{(3)}  &=& -16 \pi G m_2   u_1^-{}_\beta \frac{\partial}{\partial b_0^\mu } \iint   d\tau d\tau' \int \frac{d^4 k}{(2\pi)^4} \frac{ e^{ik \cdot b_0 }}{k^2}  e^{ik \cdot u_1^-\tau}
e^{ -ik \cdot  u_2^-\tau'}  \nonumber\\
&=& -16 \pi G m_2   u_1^-{}_\beta \frac{\partial}{\partial b_0^\mu }   \int \frac{d^4 k}{(2\pi)^2} \frac{ e^{ik \cdot b_0 }}{k^2}  \delta(k \cdot u_1^-)
\delta(k \cdot  u_2^-)  \nonumber\\
&=& 8  G m_2   u_1^-{}_\beta   \frac{b_0{}_\mu}{b_0^2}   \,.
\end{eqnarray}
Also in this case ${\mathcal C}_{xy}^{(3)}  ={\mathcal C}_{yx}^{(3)}  =0$.
 
\subsection{Computation of ${\mathcal C}^{(4)}_{\mu\alpha}$}
We have
\begin{eqnarray}
{\mathcal C}_{\mu\beta}^{(4)}  &=& 16 \pi G m_2     \iint   d\tau d\tau' \int \frac{d^4 k}{(2\pi)^4} \frac{ik_\mu e^{ik \cdot b_0 }}{k^2}  e^{ik \cdot u_1^-\tau}
e^{ -ik \cdot  u_2^-\tau'}  \delta u_1{}_\beta(\tau) \nonumber\\
&=& 16 \pi G m_2     \int   d\tau  \delta u_1{}_\beta(\tau)  \int \frac{d^4 k}{(2\pi)^3} \frac{ik_\mu  e^{ik \cdot (b_0 +u_1^-\tau)}}{k^2}  
\delta(k \cdot  u_2^-)  \,.
\end{eqnarray}
Let 
\beq
\xi=b_0 +u_1^-\tau\,,\qquad \eta =P(u_2^-)\xi =b_0+\tau \sqrt{\gamma^2-1}e(u_2^-)_2\,.
\eeq
We find
\begin{eqnarray}
{\mathcal C}_{\mu\beta}^{(4)} &=& 16 \pi G m_2   \int   d\tau  \delta u_\beta(\tau)  \frac{\partial}{\partial \eta^\mu } \int \frac{d^4 k}{(2\pi)^3} \frac{ e^{iP(u_2^-)k \cdot (b_0 +u_1^-\tau)}}{|P(u_2^-)k|^2} 
\frac{1}{\gamma}\delta(k^0+\gamma^{-1}\sqrt{\gamma^2-1}k^y)  \nonumber\\
&=&  16 \pi G m_2    \int   d\tau  \delta u_1{}_\beta(\tau) \frac{\partial}{\partial \eta{}^\mu} \int \frac{d k^x d(\gamma^{-1} k^y)dk^z}{(2\pi)^3} \frac{ e^{i(k^xb +\frac{\sqrt{\gamma^2-1}}{\gamma} k^y \tau)}}{(k^x)^2+(\gamma^{-1}k^y)^2+(k^z)^2} 
 \nonumber\\
&=& 4 G m_2    \int   d\tau \delta u_1{}_\beta(\tau)  \frac{\partial}{\partial \eta^\mu} \frac{1}{\sqrt{b_0^2+(\gamma^2-1)\tau^2}}   \nonumber\\
&=& -4 G m_2   \int   d\tau \delta u_1{}_\beta(\tau)   \frac{\eta_\mu}{[b_0^2+(\gamma^2-1)\tau^2]^{3/2}}   \nonumber\\
&=& {\mathcal C}_{\mu x}^{(4)} \delta_\beta^x + {\mathcal C}_{\mu y}^{(4)} \delta_\beta^y \,.
\end{eqnarray}
In particular
\beq
{\mathcal C}_{x y}^{(4)}=2 G^2 m_2{}^2 \pi  \frac{\gamma(2\gamma^2-3) }{ \gamma^2-1 }\frac{1}{b_0^2}\,,\qquad {\mathcal C}_{yx}^{(4)}=2G^2 m_2{}^2\pi \frac{\gamma (2\gamma^2-1)}{(\gamma^2-1)}\frac{1}{b_0^2}\,.
\eeq

We can now proceed summing up all the various contributions 
(generated by the the first order in ${\sf h}$ terms of the metric) entering $\Lambda_1^1{}_2$, i.e., recalling Eq. \eqref{eqA10}  
\beq
\Lambda_{(1)}^1{}_2=\frac12 ( {\mathcal B}_{xy}-{\mathcal B}_{yx})-\frac14 ( {\mathcal C}_{xy}-{\mathcal C}_{yx})= {\mathcal B}_{[xy]}-\frac12 {\mathcal C}_{[xy]}\equiv 
\Lambda_{{\mathcal B}}{}^x{}_y+\Lambda_{{\mathcal C}}{}^x{}_y
\,,
\eeq
with $\Lambda_{{\mathcal B}}{}^x{}_y=\sum_{n=1}^6 {\mathcal B}^{(n)}_{[xy]}$ and $\Lambda_{{\mathcal C}}{}^x{}_y=-\frac12 \sum_{n=1}^4 {\mathcal C}^{(n)}_{[xy]}$.
Summing up the various contributions to $\Lambda_{{\mathcal B}}{}^x{}_y$ we find:
\begin{align}
{\mathcal B}^{(1)}_{[xy]} &=  +12 G m_2 \frac{\gamma}{b_0} & +0\cr 
{\mathcal B}^{(2)}_{[xy]} &= 0  &  - \pi G^2 m_1 m_2 \frac{\gamma (2\gamma^2-3)}{b_0^2}\cr
{\mathcal B}^{(3)}_{[xy]} &= 0  & - \pi G^2  m_2{}^2 \frac{\gamma (2\gamma^2-3)}{b_0^2} \cr
{\mathcal B}^{(4)}_{[xy]} &=-4   G m_2 \frac{\gamma}{b_0} &  -2 G^2  m_2{}^2 \left(-\frac12 \pi(2\gamma^2+1) +\frac{2(2\gamma^2-1)}{\gamma^2-1} \right)\frac{\gamma }{b_0^2}  \cr
{\mathcal B}^{(5)}_{[xy]} &=-4 G m_2 \frac{\gamma}{b_0} &  -2 G^2  m_1 m_2 \left(-\frac12 \pi(2\gamma^2+1) +\frac{2(2\gamma^2-1)}{\gamma^2-1} \right)\frac{\gamma }{b_0^2} \cr
{\mathcal B}^{(6)}_{[xy]} &=0  & - \pi G^2 m_1 m_2 \frac{\gamma^3(2\gamma^2-3)}{\gamma^2-1}\frac{1}{b_0^2} + \pi G^2 m_1 m_2    \frac{(2\gamma^2-1)}{\gamma^2-1}\frac{\gamma}{b_0^2}
\end{align}
Similarly, summing the various contributions to $-2\Lambda_{{\mathcal C}}{}^x{}_y=\sum_{n=1}^4 {\mathcal C}^{(n)}_{[xy]}$ we find:
\begin{align}
{\mathcal C}^{(1)}_{[xy]} &=0  & +0  \cr 
{\mathcal C}^{(2)}_{[xy]} &=0  & +0  \cr
{\mathcal C}^{(3)}_{[xy]} &=0  & +0  \cr
{\mathcal C}^{(4)}_{[xy]} &=0  & + \pi  G^2 m_2{}^2  \frac{\gamma(2\gamma^2-3) }{ \gamma^2-1 }\frac{1}{b_0^2} - G^2 m_2{}^2\pi \frac{\gamma (2\gamma^2-1)}{(\gamma^2-1)}\frac{1}{b_0^2}  
\end{align}
Including the quadratic term
\beq
\int_{-\infty}^\infty d\tau \int_{-\infty}^\tau d\tau' \omega_1(\tau)^x{}_\beta  \omega_1(\tau')^\beta{}_y = -\frac{\pi G^2 m_2{}^2}{2}\frac{\gamma (4\gamma^2-3)}{b_0^2 (\gamma^2-1)}
\eeq
and comparing with our previous computations (those using only the 1PM metric), namely  
\beq
\begin{aligned}
\Lambda_{(1)}^1{}_2&=\frac{4Gm'\gamma}{b_0}+\frac{\gamma G^2 m_2{}^2(2\gamma^2+1)\pi}{b_0^2}-4G^2m'{}^2\gamma \frac{(2\gamma^2-1)}{b_0^2(\gamma^2-1)}  \qquad\qquad & [h_1]\cr
&- 2G^2 m_1 m_2\pi \frac{\gamma (\gamma^2-2)(2\gamma^2-1)}{b_0^2(\gamma^2-1)} \qquad\qquad &[\delta u_2]\cr
& +G^2 m_1 m_2 \pi \frac{\gamma (2\gamma^2+1)}{b_0^2}-4 G^2 mm' \frac{\gamma (2\gamma^2-1)}{b_0^2(\gamma^2-1)}  \qquad\qquad  & [\delta r_2] \cr
& +G^2 m_2{}^2\pi \frac{\gamma (2\gamma^2-1)}{2b_0^2 (\gamma^2-1)}   \qquad\qquad  & [\delta u_1^x] \cr
& -G^2 m_2{}^2\pi \frac{\gamma (2\gamma^2-1)(2\gamma^2-3)}{2b_0^2 (\gamma^2-1)}   \qquad\qquad  & [\delta u_1^y] \cr
&  -\frac{G^2 m_2{}^2\pi }{2}\frac{\gamma (4\gamma^2-3)}{b_0^2 (\gamma^2-1)} \qquad \qquad & [(\omega_1)^2]
\end{aligned}
\eeq
we find complete agreement.

\end{widetext}

\section{The impact parameter at 1PM order}

The explicit calculation showing the relation between $b$ and $b_0$ uses
\beq
L_\lambda=U^{\rm as}{}^\sigma \epsilon_{\sigma\mu\nu \lambda} (m_1z_1^\mu(\tau)u_1^\nu(\tau)+m_2z_2{}^\mu(\tau')u_2{}^\nu(\tau'))
\eeq
where  $U^{\rm as}$ has been defined in Eq. \eqref{Uas_def}.
We find $L_\lambda=L \delta_\lambda^z=L_0 +G\delta L$ with
\begin{eqnarray}
\label{L_def_form}
L_0 &=& m b \sinh \alpha \nonumber\\
\delta L &=&  \left[m_2\sqrt{\gamma^2-1}(\delta u_2{}^x\tau'-\delta z_2{}^x(\tau'))\right. \nonumber\\
&& \left. +m_1b \delta u_1^y(\tau)  \right]\cosh \alpha  \nonumber\\
&&  
- \left[ m_2\gamma (\delta u_2{}^x\tau'-\delta z_2{}^x(\tau'))\right.\nonumber\\
&& \left. +m_1 (\delta u_1^x \tau-\delta z_1^x(\tau) ) \right]\sinh \alpha \,.
\end{eqnarray}

In this expression $\tau$ must be replaced by its functional link with $\tau'$ implied by the definition of center-of-mass frame:
\beq
U^{\rm as}\cdot (z_2(\tau')-z_1(\tau))=0\,,
\eeq
that is
\beq
\tau =\left(\gamma_0-\sqrt{\gamma^2-1}\tanh \alpha \right) \tau' + G \delta \tau'\,.
\eeq
We see that it is enough to know the relation $\tau$  vs $\tau'$  at the zeroth order in $G$, which implies
\begin{eqnarray}
\delta L 
&=&m_1 \sinh \alpha [(\delta u_2{}^x(\tau')\tau'-\delta z_2{}^x(\tau'))\nonumber\\
&& -(\delta u_1^x(\tau) \tau-\delta z_1^x(\tau) )]\nonumber\\
&&  + m_1 b \cosh \alpha\delta u_1^y(\tau)\,.
\end{eqnarray}
Taking into account that for $\tau \to +\infty$
\beq
\frac{S(\tau)}{D(\tau)}\tau -\frac{S-1}{\sqrt{\gamma^2-1}}=\frac{1}{\sqrt{\gamma^2-1}}+O\left( \frac{1}{\tau}\right) \,,
\eeq
we find
\beq
\delta u_2{}^x(\tau')\tau'-\delta z_2{}^x(\tau')=\left\{\begin{array}{ll}
+\infty & -Gm_2\frac{ 1-2\gamma^2 }{ \gamma^2-1 }\cr
 & \cr
-\infty & -Gm_1\frac{1-2\gamma^2}{\gamma^2-1}\,,
\end{array}
\right.
\eeq
and
\beq
\delta u_1{}^x(\tau)\tau-\delta z_1^x(\tau)=\left\{\begin{array}{ll}
+\infty &Gm_1 \frac{ 1-2\gamma^2 }{ \gamma^2-1 } \cr
& \cr
-\infty & +Gm_2\frac{1-2\gamma^2}{\gamma^2-1}\,.
\end{array}
\right.
\eeq
Moreover,
\beq
\lim_{\tau =\pm \infty}\delta u^y(\tau) =0\,.
\eeq
Taking the limit $\tau' \to -\infty$ gives
\begin{eqnarray}
L_- &=& m_1 b_0 \sinh \alpha + G \frac{(2\gamma^2-1)}{\gamma^2-1}m_1 (m_1+m_2) \sinh \alpha\nonumber\\
&=&m_1 \sinh \alpha (b_0 + G \delta b)\,,
\end{eqnarray}
and hence
\beq
\delta b= \frac{(2\gamma^2-1)}{\gamma^2-1} (m_1+m_2)\,.
\eeq

Taking the limit $\tau' \to +\infty$ gives exactly the same limit, i.e.,
\beq
L_+ =m_1 b_0 \sinh \alpha + G \frac{(2\gamma^2-1)}{\gamma^2-1}m_1 (m_1+m_2) \sinh \alpha \,,
\eeq
in agreement with  previous results.

\section*{Acknowledgments}
D.B. thanks the International Center for  
Relativistic Astrophysics Network (ICRANet) and the  
Italian Istituto Nazionale di Fisica Nucleare (INFN) for  
partial support and the Institut des Hautes Etudes 
Scientifiques (IHES) for warm hospitality at various stages  
during the development of the present project.

\end{document}